\documentclass[twocolumn,prb,a4paper,floatfix]{revtex4}
\usepackage{amssymb}
\usepackage{graphicx}
\usepackage{amsmath}
\usepackage{graphicx}
\usepackage{dcolumn}
\usepackage{bm}
\usepackage{color}

\begin{document}

\title{Polarization of graphene in a strong magnetic field beyond the Dirac cone approximation}
\author{Shengjun Yuan$^{1}$, Rafael Rold\'{a}n$^{2}$, and Mikhail I.
Katsnelson$^1$}
\date{\today }
\affiliation{\centerline{$^1$Institute for Molecules and Materials, Radboud University of
Nijmegen, NL-6525AJ Nijmegen, The Netherlands}\\
\centerline{$^2$Instituto de Ciencia de Materiales de Madrid, CSIC,
Cantoblanco E28049 Madrid, Spain}}

\begin{abstract}
In this paper we study the excitation spectrum of graphene in a strong
magnetic field, beyond the Dirac cone approximation. The dynamical
polarizability is obtained using a full $\pi$-band tight-binding model where
the effect of the magnetic field is accounted for by means of the Peierls
substitution. The effect of electron-electron interaction is considered
within the random phase approximation, from which we obtain the \textit{%
dressed} polarization function and the dielectric function. The range of validity of the Landau level quantization within the continuum approximation is studied, as well as the non-trivial quantization of the spectrum around the Van Hove singularity. We further
discuss the effect of disorder, which leads to a smearing of the absorption
peaks, and temperature, which activates additional inter-Landau level
transitions induced by the Fermi distribution function.
\end{abstract}

\maketitle

\section{Introduction}

One of the most remarkable features of graphene is its anomalous quantum
Hall effect (QHE), which reveals the relativistic character of the low
energy carriers in this material.\cite{NF05,ZTSK05} In fact, the linear
electronic dispersion of graphene near the neutrality point leads to a
relativistic quantization of the electrons' kinetic energy into
non-equidistant Landau levels (LL), with the presence of a zero-energy LL,
which is the characteristics spectrum for systems of massless Dirac fermions.%
\cite{K12,G11} As a consequence, the excitation spectrum and the screening
properties in graphene are different from those of a standard
two-dimensional electron gas (2DEG) with a quadratic band dispersion, as it
may be seen from the polarization and dielectric function in the two cases.%
\cite{S86,WSSG06,HS07,RGF10,PG11}

The Coulomb interaction between electrons in completely filled LLs leads to
collective excitations and to the renormalization of the electronic
properties such as the band dispersion and the Fermi velocity. These issues
have been studied both theoretically\cite{IWFB07,S07,GSC07,BM08,RFG09,S10,WL11}
and experimentally, in the framework of cyclotron resonance experiments.\cite%
{SH06,JS07,DG07,HS10,FP11} However, most of the theoretical work has been
based on the continuum Dirac cone approximation, which does not apply when
high energy inter-LL transitions are probed.\cite{PH08} In
recent experimental realization of \textquotedblleft artificial
graphene\textquotedblright,\cite{SP11} a two-dimensonal nanostructure that
consists of identical potential wells (quantum dots) arranged in a honeycomb
lattice, the lattice constant ($a\sim 130$ nm) is much larger than the one in
graphene ($a_{0}\sim 0.142~$nm). This provides a way to study graphene in the
ultra-high magnetic field limit, since a perpendicular magnetic field in
\textquotedblleft artificial graphene\textquotedblright\ correspondes to an
effective field which is $(a/a_{0})^{2}\sim 8\times 10^{5}$ times larger
than in graphene. Furthermore, the recently developed techniques of chemical doping\cite{MR10} and electrolytic
gating \cite{Dk10} have enabled doping graphene with ultrahigh carrier densities,
where the band structure is no longer Dirac-like and one should take into
account the full $\pi $-band structure including the Van Hove singularities (VHS).

In this paper, we present a complete theoretical study of the density of
states (DOS), the polarizability and dielectric function of graphene in a
strong magnetic field, calculated from a $\pi $-band tight-binding model.
The magnetic field has been introduced by means of a Peierls phase,\cite%
{L51,K12} and the effect of long range Coulomb interaction is accounted for
within the random phase approximation (RPA). Our method allows us to study
the effect of temperature, which leads to the activation of additional
inter-LL transitions. We also study the effect of disorder in the spectrum,
which leads to a smearing of the resonance peaks.

\section{Description of the Method}


In this section we summarize the method used in the numerical calculation of
the polarizability of graphene in the QHE regime.\footnote{%
For a comprehensive discussion of the polarizability of graphene at zero
magnetic field, we refer the reader to Ref. \onlinecite{YRK11}.} A monolayer
of graphene consists of two triangular sublattices of carbon atoms with an
inter-atomic distance of $a\approx 1.42$~\AA . By considering only first
neighbor hopping between the $p_{z}$ orbitals, the $\pi $-band tight-binding
Hamiltonian of a graphene layer is given by 
\begin{equation}
H=-\sum_{\langle i,j\rangle }(t_{ij}a_{i}^{\dagger }b_{j}+\mathrm{h.c.}%
)+\sum_{i}v_{i}c_{i}^{\dagger }c_{i},  \label{Eq:Hamiltonian}
\end{equation}%
where $a_{i}^{\dagger }$ ($b_{i}$) creates (annihilates) an electron on
sublattice A (B) of the graphene layer, and $t_{ij}$ is the nearest neighbor
hopping parameter, which oscillates around its mean value $t\approx 3$~eV.%
\cite{CG09} The second term of $H$ accounts for the effect of an on-site
potential $v_{i}$, where $n_{i}=c_{i}^{\dagger }c_{i}$ is the occupation
number operator. For simplicity, we omit the spin degree of freedom in Eq.~(%
\ref{Eq:Hamiltonian}), which contributes only through a degeneracy factor.
In our numerical calculations, we use periodic boundary conditions. The
effect of a perpendicular magnetic field $\mathbf{B}=B{\hat{\mathbf{z}}}$ is
accounted by means of the Peierls substitution, which transforms the hopping
parameters according to\cite{L51} 
\begin{equation}
t_{ij}\rightarrow t_{ij}\exp {\left( i\frac{2\pi }{\Phi _{0}}\int_{\mathbf{R}%
_{i}}^{\mathbf{R}_{j}}\mathbf{A}\cdot d\mathbf{l}\right) },
\end{equation}%
where $\Phi _{0}=hc/e$ is the flux quantum and $\mathbf{A}$ is the vector
potential, e. g. in the Landau gauge $\mathbf{A}=(-By,0,0)$. We will
calculate the DOS and the polarization function of the system by using an
algorithm based on the evolution of the time-dependent Schr\"{o}dinger
equation. For this we will use a random superposition of all basis states as
an initial state $|\varphi \rangle $ (see e. g. Refs.\onlinecite{HR00,YRK10}%
) 
\begin{equation}
\left\vert \varphi \right\rangle =\sum_{i}a_{i}c_{i}^{\dagger
}c_{i}\left\vert 0\right\rangle ,  \label{Eq:phi0}
\end{equation}%
where $a_{i}$ are random complex numbers normalized as $\sum_{i}\left\vert
a_{i}\right\vert ^{2}=1$. The DOS, which describes the number of states at a
given energy level, is then calculated as a Fourier transform of the
time-dependent correlation functions 
\begin{equation}
d(\epsilon )=\frac{1}{2\pi }\int_{-\infty }^{\infty }e^{i\epsilon \tau
}\langle \varphi |e^{-iH\tau }|\varphi \rangle d\tau ,  \label{Eq:DOS}
\end{equation}%
with the same initial state defined in Eq. (\ref{Eq:phi0}). The dynamical
polarization function can be obtained from the Kubo formula\cite{K57} 
\begin{equation}
\Pi \left( \mathbf{q},\omega \right) =\frac{i}{V}\int_{0}^{\infty }d\tau
e^{i\omega \tau }\left\langle \left[ \rho \left( \mathbf{q},\tau \right)
,\rho \left( -\mathbf{q},0\right) \right] \right\rangle ,  \label{Eq:Kubo}
\end{equation}%
where $V$ denotes the volume (or area in 2D) of the unit cell, $\rho \left( 
\mathbf{q}\right) $ is the density operator given by%
\begin{equation}
\rho \left( \mathbf{q}\right) =\sum_{i}c_{i}^{\dagger }c_{i}\exp \left( i%
\mathbf{q\cdot r}_{i}\right) ,
\end{equation}%
and the average is taken over the canonical ensemble. For the case of the
single-particle Hamiltonian, Eq.~(\ref{Eq:Kubo}) can be written as\cite%
{YRK10}%
\begin{eqnarray}
&&\Pi \left( \mathbf{q},\omega \right) =-\frac{2}{V}\int_{0}^{\infty }d\tau
e^{i\omega \tau }  \notag  \label{Eq:Kubo2} \\
&&\times \text{Im}\left\langle \varphi \right\vert n_{F}\left( H\right)
e^{iH\tau }\rho \left( \mathbf{q}\right) e^{-iH\tau }\left[ 1-n_{F}\left(
H\right) \right] \rho \left( -\mathbf{q}\right) \left\vert \varphi
\right\rangle ,  \notag \\
&&
\end{eqnarray}%
where 
\begin{equation}
n_{F}\left( H\right) =\frac{1}{e^{\beta \left( H-\mu \right) }+1}
\label{Eq:Fermi-Dirac}
\end{equation}%
is the Fermi-Dirac distribution operator, $\beta =1/k_{B}T$ where $T$ is the
temperature and $k_{B}$ is the Boltzmann constant, and $\mu $ is the
chemical potential. In the numerical simulations, we use units such that $%
\hbar =1$, and the average in Eq.~(\ref{Eq:Kubo2}) is performed over the
random superposition Eq. (\ref{Eq:phi0}). The Fermi-Dirac distribution
operator $n_{F}\left( H\right) $ and the time evolution operator $e^{-iH\tau
}$ can be obtained by the standard Chebyshev polynomial decomposition.\cite%
{YRK10}

\begin{figure}[t]
\begin{center}
\mbox{
\includegraphics[width=7.cm]{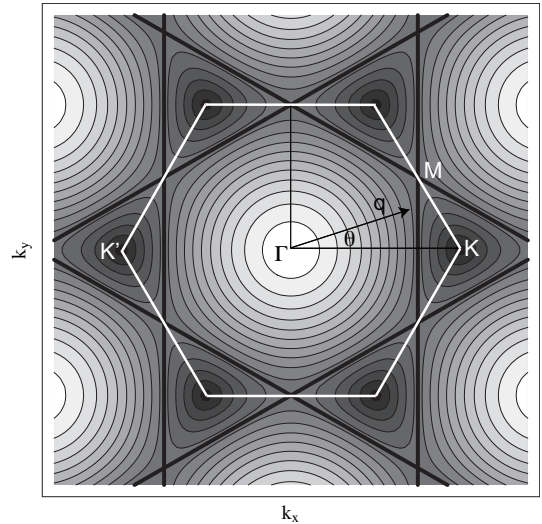}
}
\end{center}
\caption{Constant energy contours obtained from the band dispersion Eq. (%
\protect\ref{Eq:EnergyB=0}). The thick black lines correspond to dispersion
at the VHS $|\protect\epsilon| = t$. Notice that the CEC are centered around
the Dirac points for $|\protect\epsilon|<t$ and around $\Gamma$ for $|\protect%
\epsilon|>t$. For illustrative reasons, the hexagonal BZ is shown in white. For undoped graphene, the valence and
 conduction bands touch each other at the vertices of the hexagon, the so
 called Dirac points (K and K'). The Van Hove singularity lies at the M
 point, and we have defined $\protect\theta$ as the angle between the
 wave-vector $q$ and the $k_x$-axis.}
\label{Fig:FS}
\end{figure}

\begin{figure*}[t]
\begin{center}
\mbox{
\includegraphics[width=6.3cm]{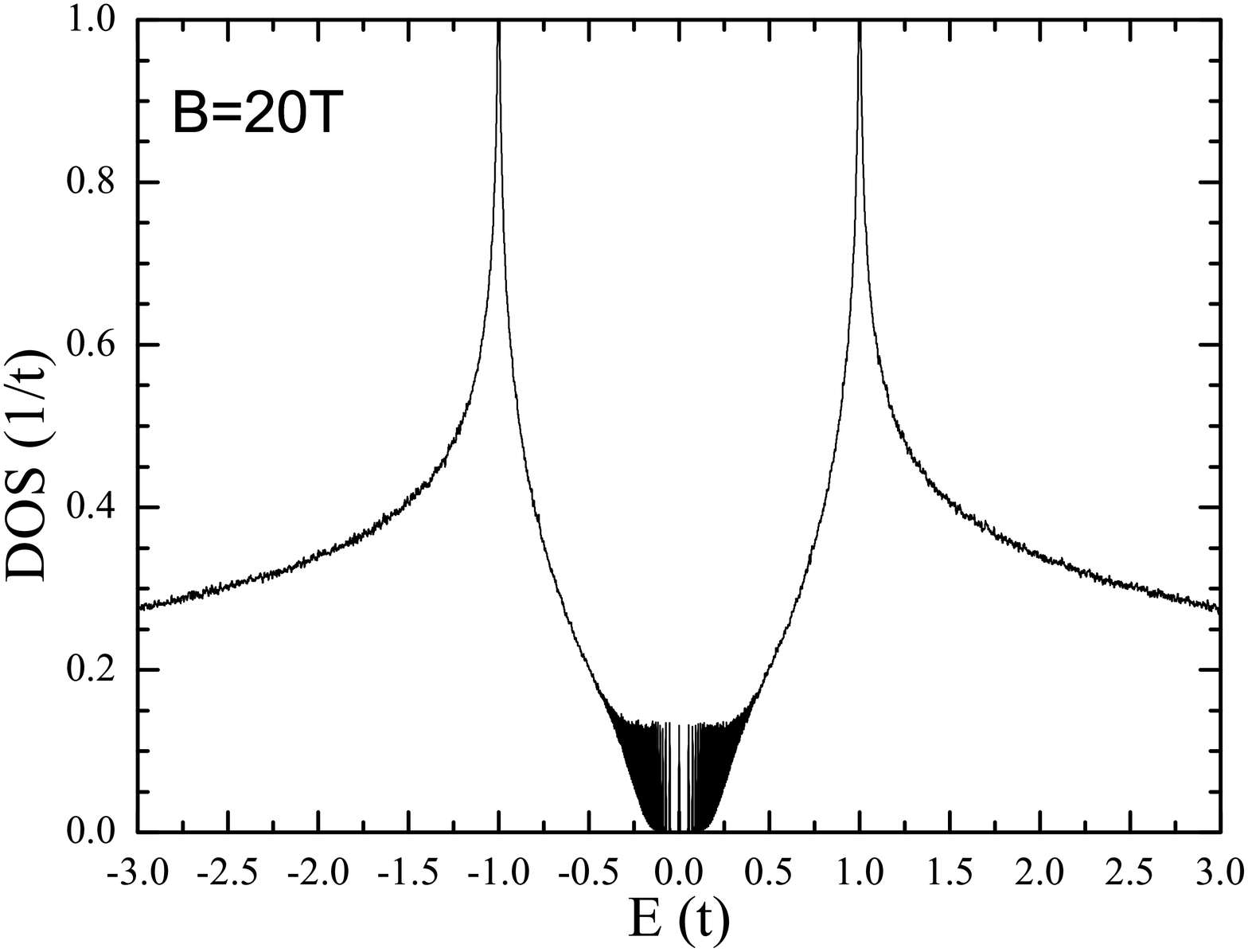}
\includegraphics[width=6.3cm]{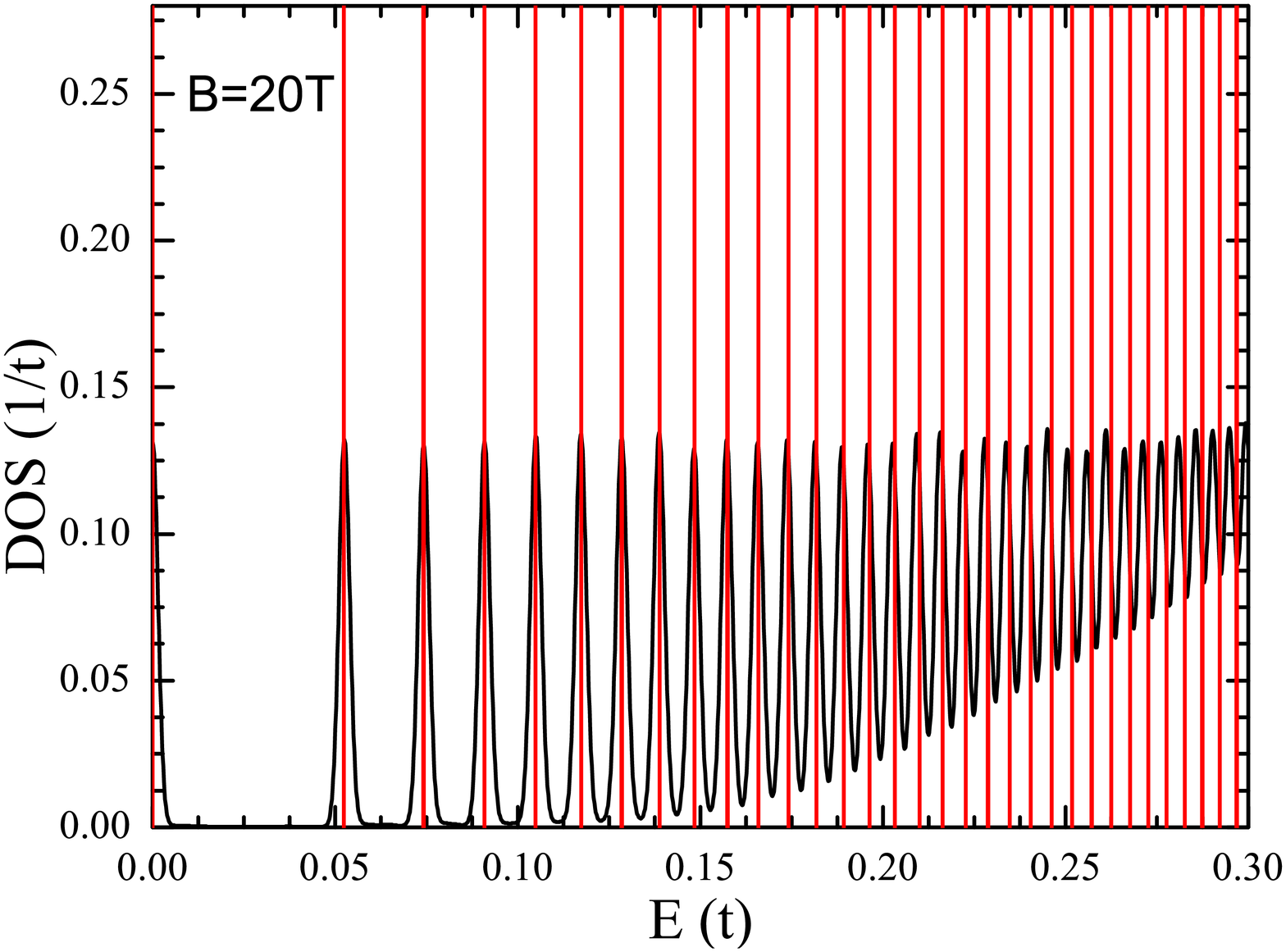}
} 
\mbox{
\includegraphics[width=6.3cm]{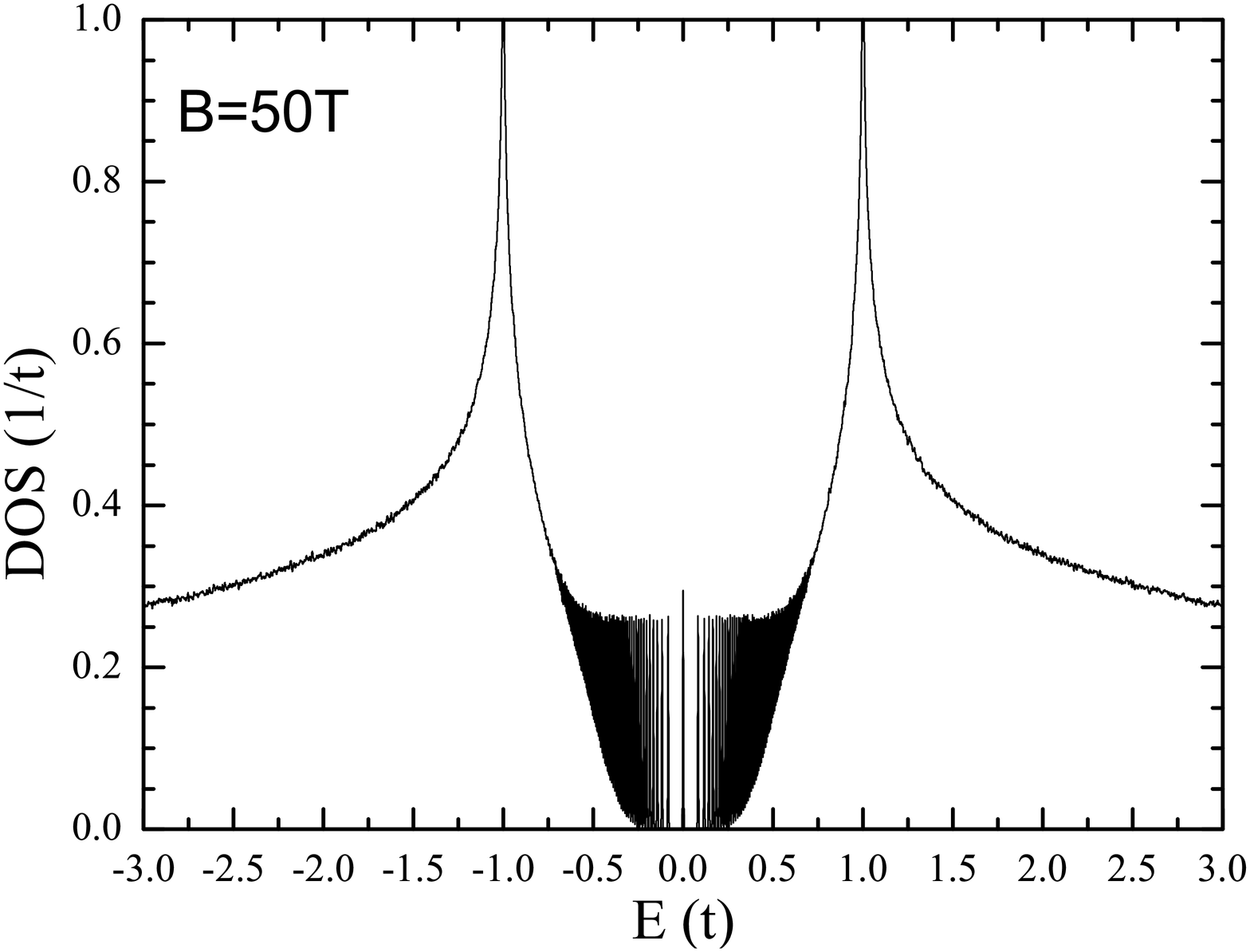}
\includegraphics[width=6.3cm]{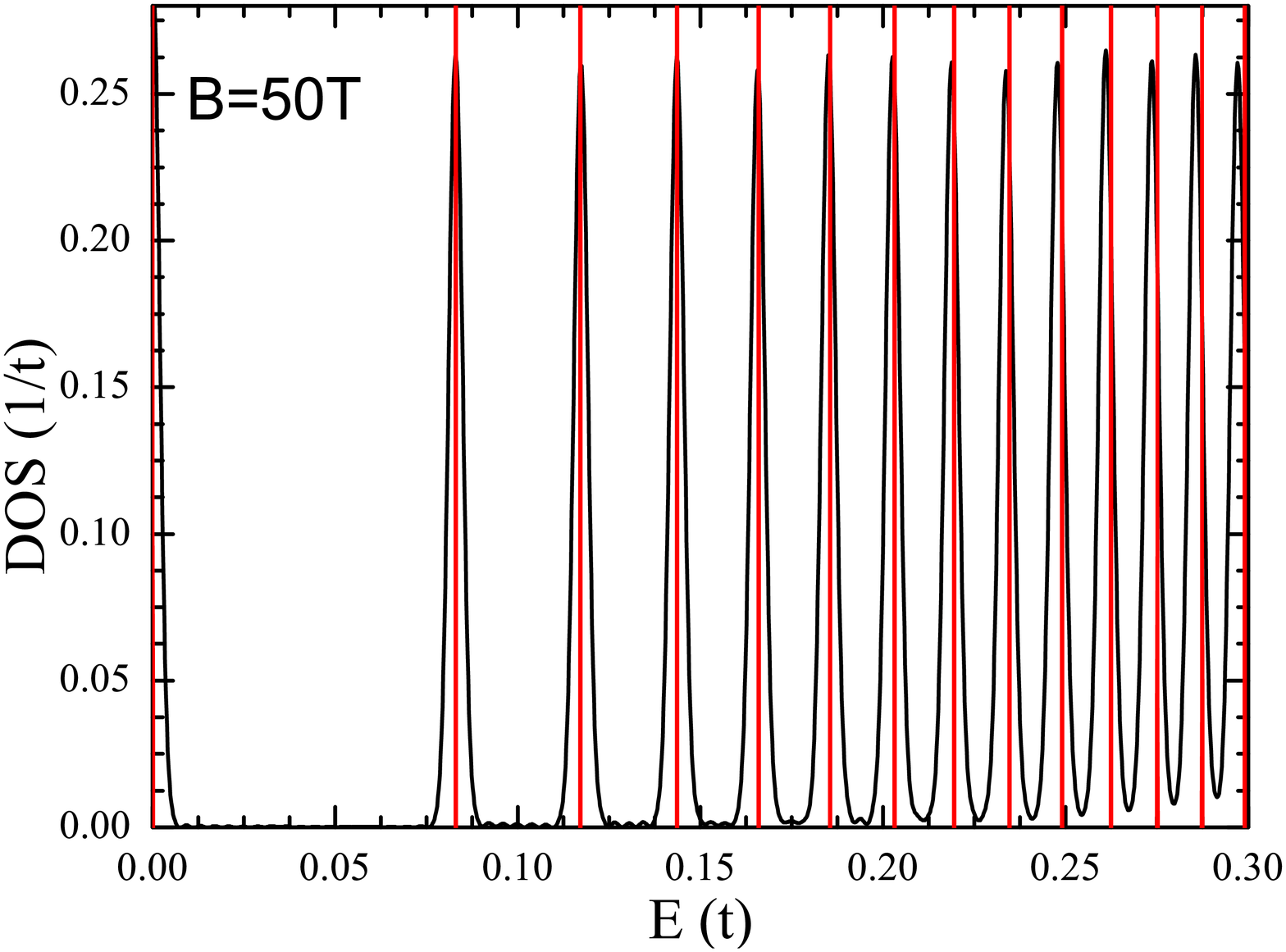}
}
\end{center}
\caption{Left: DOS of a monolayer of graphene in a magnetic field, for two
different values of $B$. Right: zoom of the low energy region of the
spectrum. The red vertical lines indicate the position of the LLs in the
continuum Dirac cone approximation, Eq. (\protect\ref{Eq:DOSDirac}). We have
used a sample made of 4096$\times $4096 atoms.}
\label{Fig:DOS}
\end{figure*}

Long range Coulomb interaction is considered in the RPA, leading to a
dressed particle-hole polarization 
\begin{equation}
\chi \left( \mathbf{q},\omega \right) =\frac{\Pi \left( \mathbf{q},\omega
\right) }{\mathbf{1}-V\left( q\right) \Pi \left( \mathbf{q},\omega \right) },
\label{Eq:chi}
\end{equation}%
where 
\begin{equation}
V\left( q\right) =\frac{2\pi e^{2}}{\kappa q}
\end{equation}%
is the Fourier component of the Coulomb interaction in two dimensions, in
terms of the background dielectric constant $\kappa $. Furthermore, the
dielectric function of the system is calculated as 
\begin{equation}
\mathbf{\varepsilon }\left( \mathbf{q},\omega \right) =1-V\left( q\right)
\Pi \left( \mathbf{q},\omega \right) .  \label{Eq:epsilon}
\end{equation}%
The collective modes lead to zeroes of $\varepsilon (\mathbf{q},\omega )$,
and their dispersion relation is defined from 
\begin{equation}
\mathrm{Re}~\varepsilon (\mathbf{q},\omega _{pl})=1-V(q)\Pi (\mathbf{q}%
,\omega _{pl})=0,  \label{Eq:Plasmons}
\end{equation}%
which leads to poles in the response function (\ref{Eq:chi}). The
technicalities about the accuracy of the numerical results have been
discussed elsewhere.\cite{HR00,YRK10,YRK11} Here we just mention that the
efficiency of the method is mainly determined by three factors: the time
interval of the propagation, the total number of time steps, and the size of
the sample. The method is more efficient in the
presence of strong magnetic fields. Because for weak fields (e.g., $B<1$~T) the energy difference between LLs become very
small, this makes that the total number of time steps and the size of the sample have to
be large enough to provide the necessary energy resolution in the numerical
simulation.

\section{Density of states and excitation spectrum}

In this section we study the DOS and the excitation spectrum of a graphene
layer in a magnetic field, neglecting the effect of disorder and
electron-electron interaction. The $B=0$ dispersion relation of the $\pi $
bands obtained from a tight-binding model with nearest-neighbor hopping
between the $p_{z}$ orbitals is 
\begin{equation}
\epsilon (\mathbf{k})=\lambda t|\phi _{\mathbf{k}}|  \label{Eq:EnergyB=0}
\end{equation}%
where $\lambda =\pm 1$ is the band index and 
\begin{equation}
\phi _{\mathbf{k}}=1+2e^{i3k_{x}a/2}\cos \left( \frac{\sqrt{3}}{2}%
k_{y}a\right) .
\end{equation}%
The band dispersion Eq. (\ref{Eq:EnergyB=0}) consists of two bands that
touch each other in the vertices of the hexagonal Brilloin zone (BZ) (Fig. %
\ref{Fig:FS}), which are the so called Dirac points. In the absence of
longer range hopping terms, the band structure is electron-hole symmetric,
and the constant energy contours (CEC) obtained from Eq. (\ref{Eq:EnergyB=0}%
) are shown in Fig. \ref{Fig:FS}. For undoped graphene ($\mu =0$), which is
the band filling that we will consider all along this paper, the Fermi
surface consists of just six points at the vertices of the BZ. In this case,
the low energy excitations can be described by means of an effective theory
obtained from an expansion of the dispersion Eq. (\ref{Eq:EnergyB=0}) around
the K points. This leads to an approximate dispersion $\epsilon (\mathbf{k}%
)\approx \lambda v_{F}k$, where $v_{F}=3ta/2$ is the Fermi velocity.

If we now consider the effect of a perpendicular magnetic field,
the Landau quantization of the kinetic energy leads to a set of LLs, which
can be described from the semiclassical condition\cite{S84,K12}
\begin{equation}
S(C)=\frac{2\pi}{l_B^2}\left(n+\frac{1}{2}-\frac{\Gamma(C)}{2\pi}\right)
\end{equation}
where 
\begin{equation}
S(C)=\underset{\epsilon \left(
k_{x},k_{y}\right) \leq \text{ }\epsilon _{n}}{\int \int }dk_{x}dk_{y} 
\end{equation}
is the area enclosed by the cyclotron orbit $C$ in momentum space [for circular orbits $S(C)$ is just $\pi k^2$], $l_{B}=\sqrt{\hbar c/eB}$ is the magnetic length, $n$ is the LL index and $\Gamma(C)$ is the Berry phase. In graphene, as we will discuss in the next section, $\Gamma(C)=\pi$ for orbits around the K and K' points, and 0 for orbits around the $\Gamma$ point.\cite{FM10} From the energy dependence of $S(C)$ one can calculate the energy of the Landau levels as
\begin{equation}\label{Eq:SemiClasLLs}
\epsilon_n=S^{-1}\left( \frac{2\pi}{l_B^2}\left[n+\frac{1}{2}-\frac{\Gamma(C)}{2\pi}\right]\right),
\end{equation}
where $S^{-1}(x)$ is the inverse function to $S(x)$. Using Eq. (\ref{Eq:SemiClasLLs}), it is easy to check that the LL quantization corresponding to a low energy parabolic band $\epsilon(k)=k^2/2m_b$ (where $m_b$ is the effective mass) with a $\Gamma(C)=0$ Berry phase, is $\epsilon_n=\omega_c(n+1/2)$, where $\omega_c=eB/m_b$ is the cyclotron frequency. On the other hand, a linearly dispersing band as the one for graphene leads to a LL quantization around the Dirac points as
\begin{equation}
\epsilon _{\lambda ,n}=\lambda \epsilon _{n}=\lambda \frac{v_{F}}{l_{B}}\sqrt{2n}%
\propto \sqrt{Bn}.
\end{equation}

\begin{figure}[t]
\begin{center}
\includegraphics[width=8.5cm]{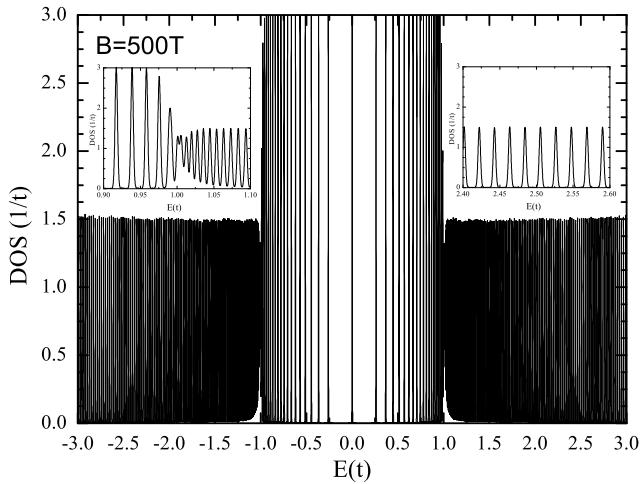}
\end{center}
\caption{DOS obtained from Eq. (\protect\ref{Eq:DOS}) at $B=500$T. The
insets show a zoom of the DOS around the VHS $|\protect\epsilon|\approx t$
and at $|\protect\epsilon| \approx 2.5t$. }
\label{Fig:DOS500T}
\end{figure}

\subsection{Density of states}

\label{Sec:DOS} 

The DOS close to the Dirac point can be approximated by\cite%
{CG09} 
\begin{equation}  \label{Eq:DOSDirac}
d_{\mathrm{Dirac}}(\epsilon)\approx\frac{2A_c}{\pi}\frac{|\epsilon|}{v_F^2}
\end{equation}
where $A_c=3\sqrt{3}a^2/2$ is the unit cell area. In Fig. \ref{Fig:DOS} we
show the DOS for two different values of the magnetic field. The black line
corresponds to the numerical tight-binding result obtained from Eq. (\ref%
{Eq:DOS}). Near $\epsilon=0$ we notice the presence of a zero energy LL
surrounded by a set of LLs whose separation decreases as the energy
increases, leading to a stacking of the LLs as we move away from the Dirac
points. The presence of a finite broadening in these LLs is
due to the energy resolution of the numerical simulations, which is limited
by the number of atoms used in the calculation, as well as the total number
of time steps, which determines the accuracy of the energy eigenvalues. In
order to check the range of validity of the continuum approximation, in the
right-hand side of Fig. \ref{Fig:DOS} we show a zoom of the positive low
energy part of the DOS for the two values of $B$, comparing the DOS obtained
with the full $\pi$-band tight-binding model Eq. (\ref{Eq:DOS}) [black
lines] to the Dirac cone approximation of Eq. (\ref{Eq:DOSDirac}) [vertical
red lines]. Contrary to multi-layer graphenes, for which trigonal warping
effects are important at rather low energies,\cite{YRK11b} we see that the
deviations of the LL positions in the continuum approximation Eq. (\ref%
{Eq:DOSDirac}) with respect to the full $\pi$-band model are weak even at
energies of the order of $\epsilon \sim 0.3t \sim 1$eV, in agreement with
magneto-optical transmission spectroscopy experiments.\cite{PH08}

A much less investigated issue is the effect of the magnetic field on the
DOS around the VHS $|\epsilon |\approx t$. For illustrative reasons we show
in Fig. \ref{Fig:DOS500T} the numerical results for the DOS of a graphene
layer at an extremely high magnetic field of $B=500$T.\footnote{%
Although this situation is unrealistic for a graphene membrane, the results
can be useful to better understand the Landau quantization and the
collective modes of artificially created honeycomb lattices where the large
value of the lattice constant $a\sim 130$nm allows for the study of the
ultra-high magnetic field limit with $l_{B}\lesssim a$.\cite{SP11}} At this
energy the LL quantization is highly nontrivial because of the saddle point
in the band structure at which there is a transition from CECs encircling the
Dirac points, to CECs encircling the $\Gamma $ point, as it can be seen in
Fig. \ref{Fig:FS}. Because in the semiclassical limit, the cyclotron orbits
in reciprocal space follow the CECs, we have that at the saddle point there
is a change in the topological Berry phase $\Gamma (C)$ from $\Gamma (C)=\pm
\pi $ for orbits encircling the Dirac points ($|\epsilon |<t$) to $\Gamma
(C)=0$ for orbits encircling the $\Gamma $ point ($|\epsilon |>t$).\cite%
{FM10} The different character of the cyclotron orbits at both sides of the
saddle point leads to two series of LLs, with different cyclotron
frequencies $\omega _{c}=eB/m_{b}$, that \textit{merge} at the VHS, as it
may be seen in the left-hand side inset of Fig. \ref{Fig:DOS500T}. Because
of the effective mass $m_{b}$ below the VHS is larger than the one above it
(the band below the saddle point is flatter than above it), the cyclotron
frequencies are also different $\omega _{c}(|\epsilon |<t)<\omega
_{c}(|\epsilon |>t)$, and consequently the LLs are more separated for $%
|\epsilon |>t$ than for $|\epsilon |<t$. The possibility of placing the
chemical potential at the VHS would bring the chance of studying highly
anomalous inter-LL transitions, due to the different separation of the LLs
above and below the VHS. Finally, at an even higher energy, the LL
quantization is quite similar to that of a 2DEG with a parabolic dispersion,
with a set of roughly equidistant LLs, as it may be seen in the right-hand
side inset of Fig. \ref{Fig:DOS500T} for $|\epsilon |\approx 2.5t$. However,
we emphasize that for realistic values of magnetic field, the LL
quantization in graphene is inappreciable in this range of energies, and the
DOS for energies $|\epsilon |\gtrsim 0.7t$ is similar to the DOS at $B=0$,%
\cite{YRK10} as it may be seen in Fig. \ref{Fig:DOS}.


\begin{figure*}[t]
\begin{center}
\mbox{
\includegraphics[width=7.3cm]{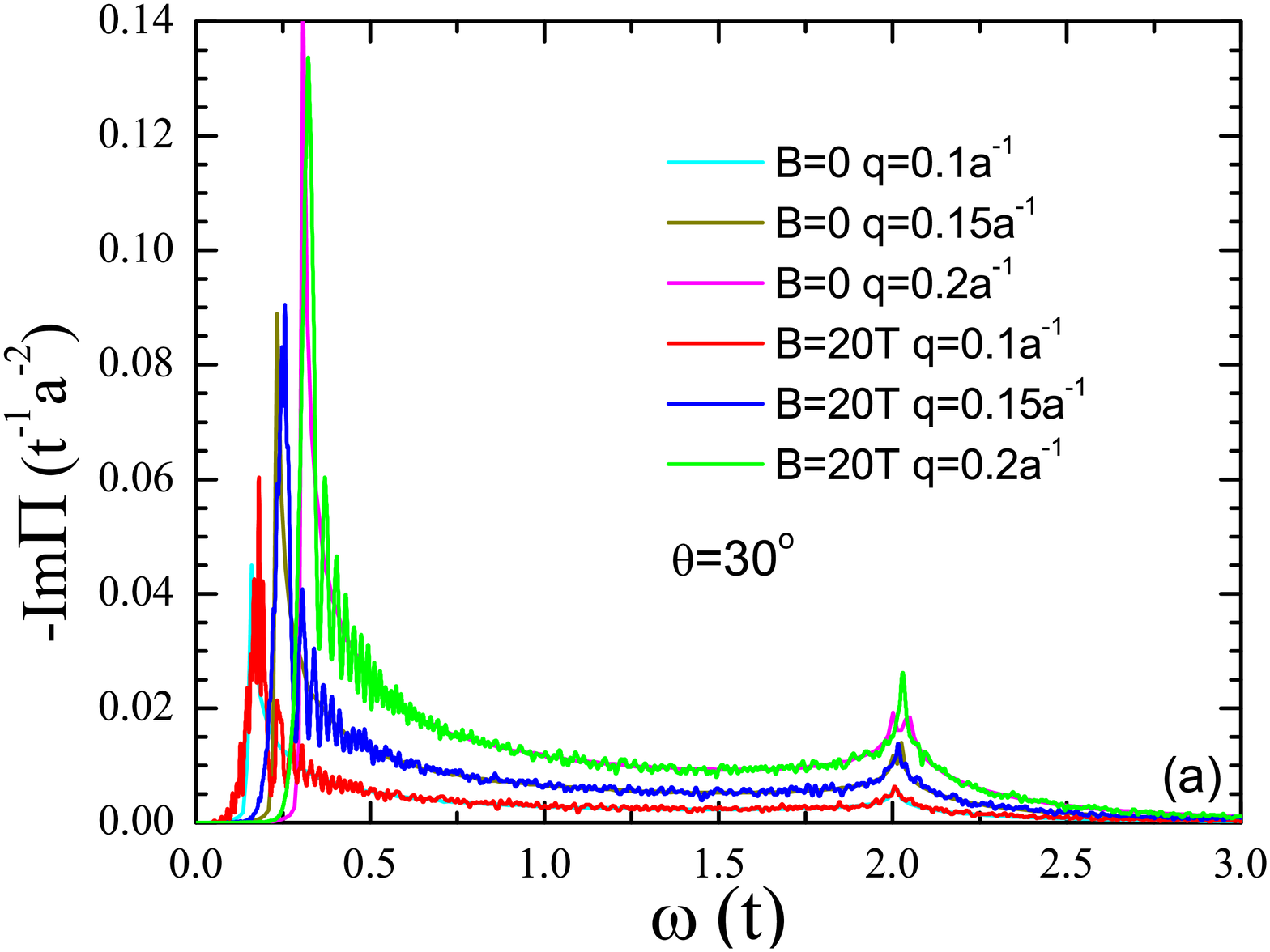}
\includegraphics[width=7.3cm]{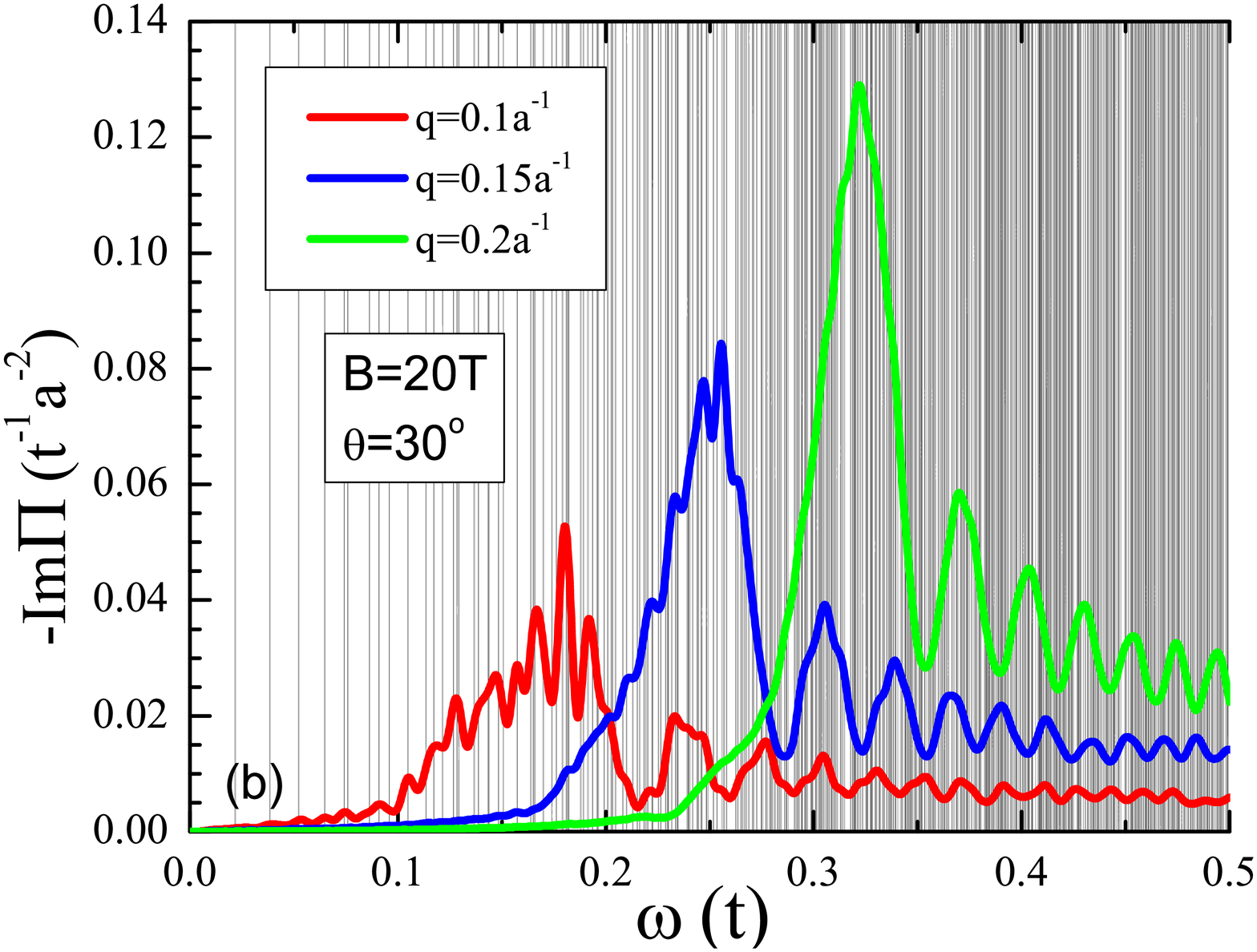}
} 
\mbox{
\includegraphics[width=7.3cm]{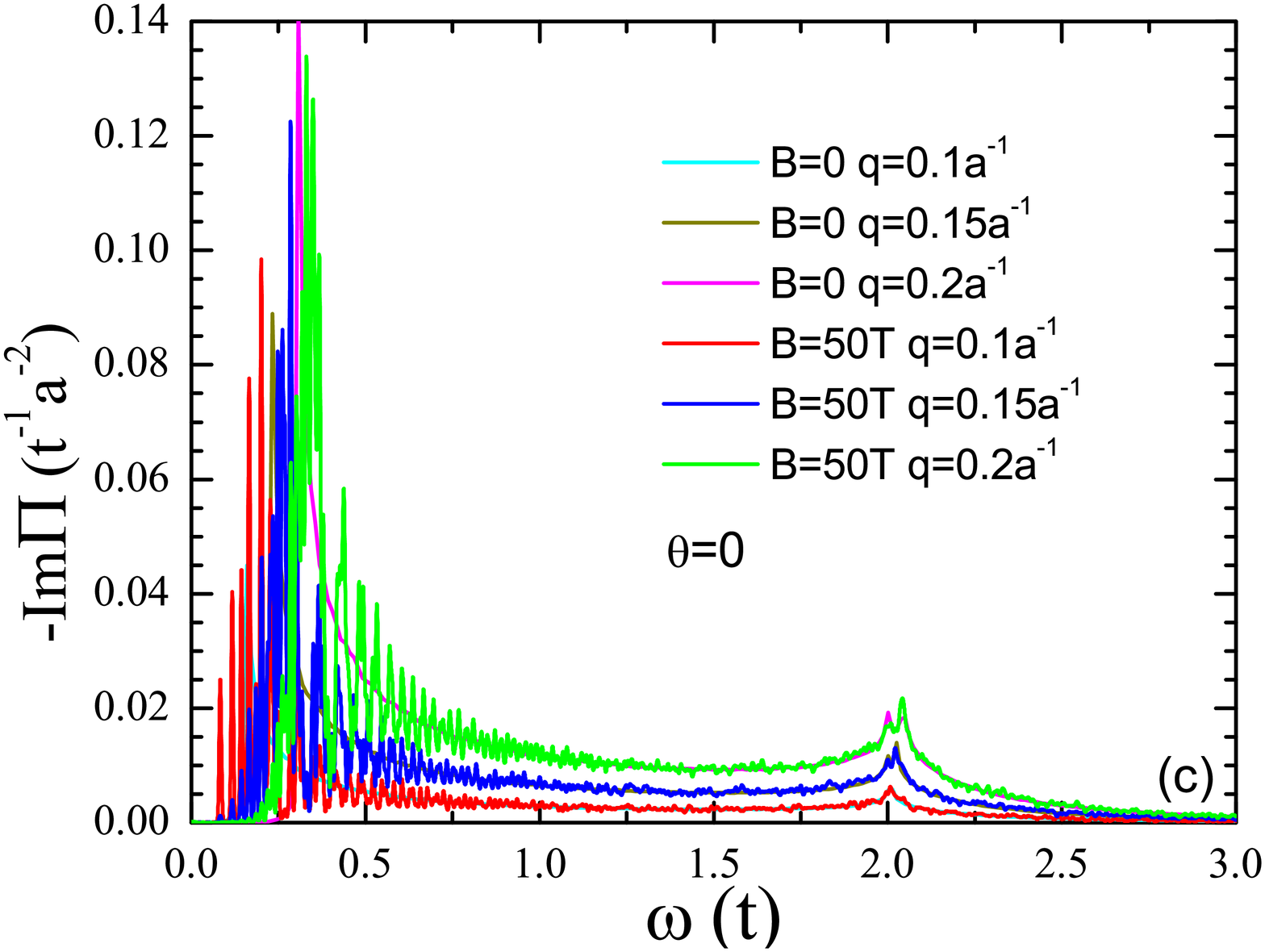}
\includegraphics[width=7.3cm]{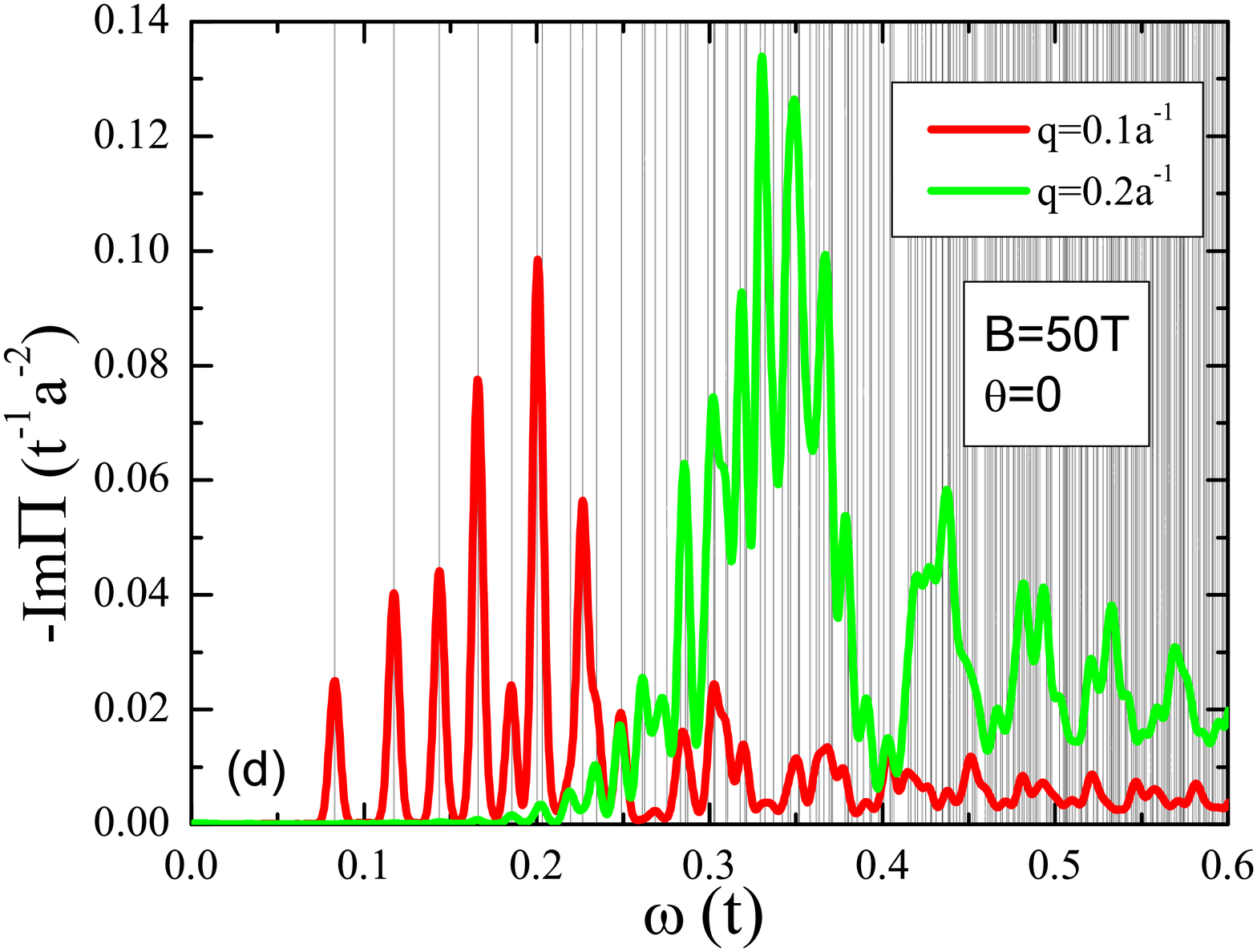}
}
\end{center}
\caption{$-\mathrm{Im}\Pi(\mathbf{q},\protect\omega)$ for different values
of wave-vector $q$ and strength of the magnetic field $B$. The angle $%
\protect\theta$ defines the orientation of the wave-vector in the Brilloin
zone (see Fig. \protect\ref{Fig:FS}). Plots (a) and (c) show the
polarization in the whole energy range. For comparison we show the
polarization at $B=0$. In plots (b) and (c) we show the low energy part of
the spectrum. The vertical black lines signal the energy of the
particle-hole processes expected from Eq. (\protect\ref{Eq:PHDirac}). }
\label{Fig:PHES}
\end{figure*}

\subsection{Particle-hole excitation spectrum}

\label{Sec:PHES} The particle-hole excitation spectrum (PHES) for
non-interacting electrons, which is the part of the $\omega-\mathbf{q}$
plane where $\mathrm{Im}\Pi(\mathbf{q},\omega)$ is non-zero, defines the
region of the energy-momentum space where particle-hole excitations are
allowed. For undoped graphene ($\mu=0$), the particle-hole excitations
correspond to inter-band transitions across the Dirac points. In Fig. \ref%
{Fig:PHES} we show $-\mathrm{Im}\Pi(\mathbf{q},\omega)$ for different values
of wave-vector and magnetic field. Two different orientations of $\mathbf{q}$
are shown, namely along the $\Gamma$-M and $\Gamma$-K directions. First, due
to the low energy linear dispersion relation and to the effect of the
chirality factor (or wave-function overlap) that suppresses backscattering,
we observe that the strongest contribution to the polarization is
concentrated around $\omega\approx v_Fq$. In fact, at $B=0$ it was shown\cite%
{GGV94} that $\mathrm{Im}\Pi(\mathbf{q},\omega)\sim
q^2(\omega^2-v_F^2q^2)^{-1/2}$, which implies an infinite response of
relativistic non-interacting electrons in graphene at the threshold $%
\omega=v_Fq$. The main difference of the PHES at finite magnetic field with
respect to its $B=0$ counterpart in this low energy range, is that in the $%
B\neq 0$ case $\Pi(\mathbf{q},\omega)$ presents a series of peaks of strong
spectral weight, due to the LL quantization of the kinetic energy, that we
will discuss in more detail below. We notice that, for the realistic values
of magnetic field used in Fig. \ref{Fig:PHES}, the spectrum at finite
magnetic field roughly coincides with the the one at $B=0$ at high energies.
This is due to the almost negligible effect of the magnetic field on the DOS
at energies $|\epsilon|\gtrsim 0.7t$ for $B\lesssim 50$T, as we saw in the
previous section. This part of the spectrum is dominated, as in the $B=0$
case,\cite{SSP10,YRK11} by a peak of $\mathrm{Im}\Pi(\mathbf{q},\omega)$
around $\omega\approx 2t$, which is due to particle-hole transitions between
states of the Van Hove singularities of the valence and the conduction bands
at $\epsilon\approx -t$ and $\epsilon\approx t$ respectively.

However, the low energy part of the spectrum is completely different to its
zero magnetic field counterpart, and it is dominated by a series of
resonance peaks at some given energy values. For the undoped case studied
here, the possible excitations correspond to inter-LL transitions of energy $%
\omega_{n,n^{\prime }}=\epsilon_{n^{\prime }}+\epsilon_{n}$, where $%
n^{\prime }$ is the LL index of the particle in the conduction band ($%
\lambda=+1$) and $n$ is the LL index of the hole in the valence band ($%
\lambda=-1$). In the continuum approximation, they have an energy 
\begin{equation}  \label{Eq:PHDirac}
\omega_{n,n^{\prime }}=\sqrt{2}(v_F/l_B)(\sqrt{n^{\prime }}+\sqrt{n}).
\end{equation}
The energy corresponding to each of these transitions is indicated by a
black vertical line in Fig. \ref{Fig:PHES}(b) and (d), where we show a zoom
of $-\mathrm{Im}\Pi(\mathbf{q},\omega)$ that corresponds to the low energy
LL transitions about the Dirac points. Notice that, in contrast to a
standard 2DEG with a parabolic dispersion and equidistant LLs, the
relativistic quantization of the energy band in graphene makes that in a
fixed energy window at high energies, there are more possible inter-LL
excitations from the level $n$ in the valence band to the level $n^{\prime }$
in the conduction band, than at low energies. As a consequence, there is a
stacking of neighboring LL transitions as we increase the energy of the
excitations, which manifests itself in a \textit{continuum} of possible
inter-LL transitions from a given energy, which for the case of Fig. \ref%
{Fig:PHES}(b) at $B=20$T is $\omega/t\gtrsim 0.25$.

Contrary to what one can naively expect, only at very strong magnetic fields
and for small values of $\omega$ and $q$, the peaks of $-\mathrm{Im}\Pi(%
\mathbf{q},\omega)$ occur at the energies given by Eq. (\ref{Eq:PHDirac}).
In fact, we can see in Fig. \ref{Fig:PHES}(d) that for $B=50$T and for the
smaller value of $q$ shown (red line), the peaks of $-\mathrm{Im}\Pi(\mathbf{%
q},\omega)$ match very well the energies for the inter-LL transitions given
by Eq. (\ref{Eq:PHDirac}). However, at weaker magnetic fields and/or larger
wave-vectors, the peaks of the polarization function do not coincide any
more with every of the inter-LL transitions given by Eq. (\ref{Eq:PHDirac}).
In fact, there is a series of peaks of $\mathrm{Im}\Pi(\mathbf{q},\omega)$,
corresponding to regions of the PHES of high spectral weight, which can be
understood from the form of the wavefunctions of the electron and the hole
that overlap to form an electron-hole pair. A detailed discussion about the
structure of the PHES in graphene in comparison with a 2DEG can be found in
Ref. \onlinecite{RGF10}. Here we just remember that the modulus of the LL
wavefunction, due to the zeros of Laguerre polynomials, presents a number of
nodes that depend on the LL index $n$. On the other hand, the existence of
an electron-hole pair will be possible if there is a finite overlap of the
electron and hole wavefunctions, which will define the form factor for
graphene in the QHE regime, $\mathcal{F}_{n,n^{\prime }}(\mathbf{q})$.
Because the node structure of the single-particle wavefunctions will be
transfered to $|\mathcal{F}_{n,n^{\prime }}(\mathbf{q})|^2$, all together
will lead to a highly modulated spectral weight in the PHES, as it is seen
in Fig. \ref{Fig:PHES}(b) and (d).

\section{Collective modes}

\begin{figure}[t]
\begin{center}
\mbox{
\includegraphics[width=7.3cm]{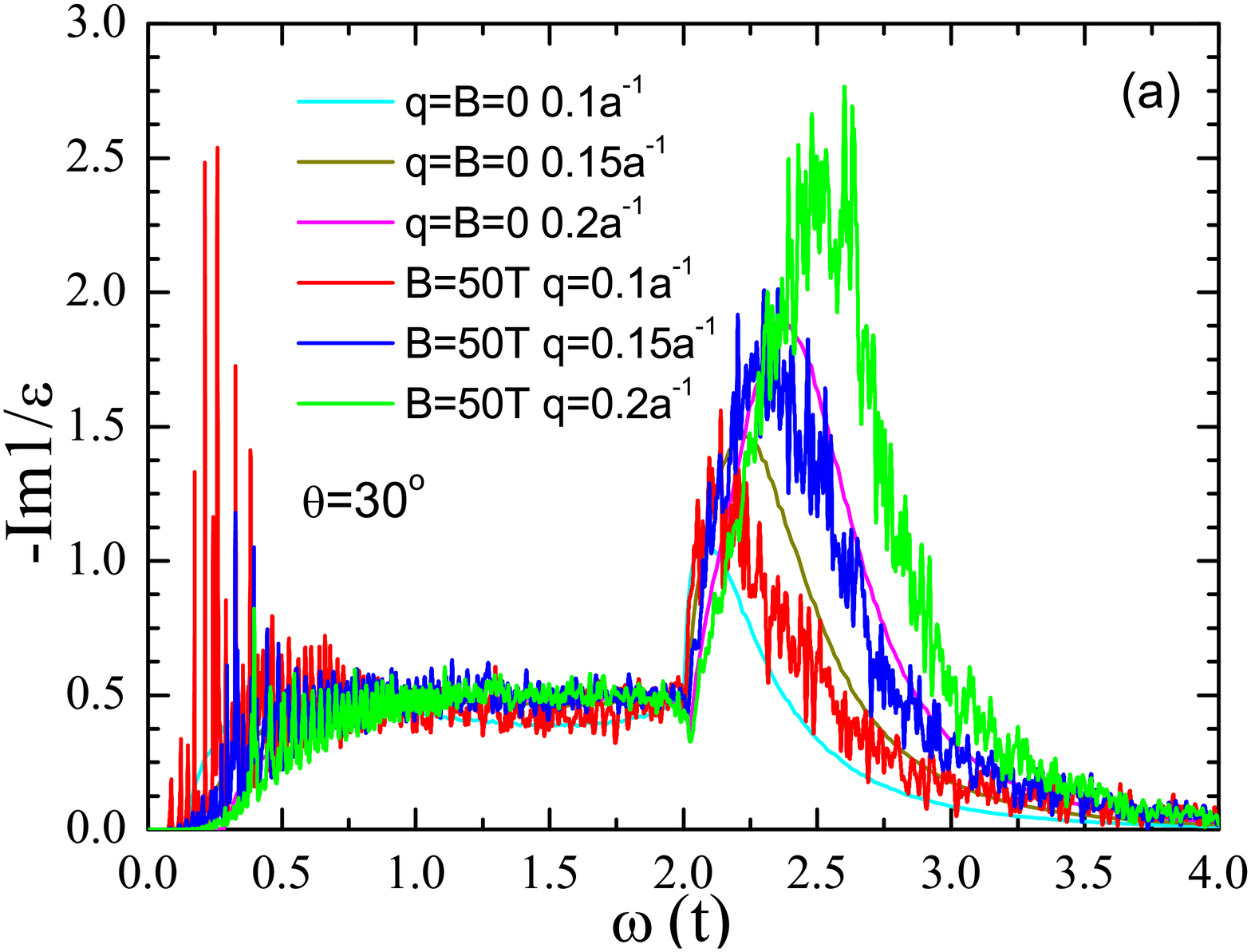}
} \mbox{
\includegraphics[width=7.3cm]{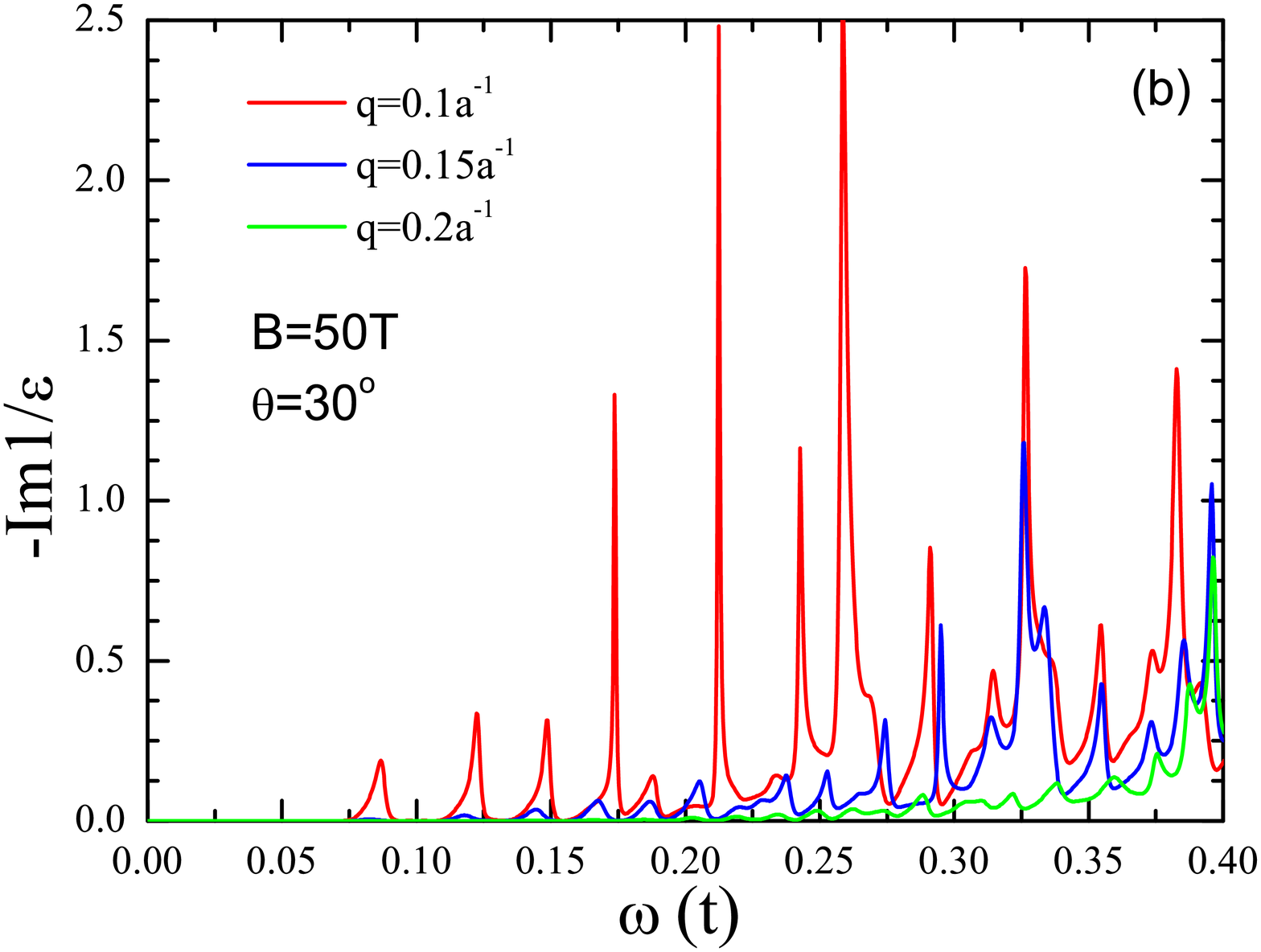}
}
\end{center}
\caption{(a) Loss function $-\mathrm{Im}~1/\protect\varepsilon(\mathbf{q},%
\protect\omega)$ in the RPA, for different values of wave-vector. The
results for graphene in a magnetic field of $B=50$T are compared to the $B=0$
case. (b) Zoom of the low energy part of the spectrum. }
\label{Fig:Loss}
\end{figure}

\begin{figure}[t]
\begin{center}
\mbox{
\includegraphics[width=4.5cm]{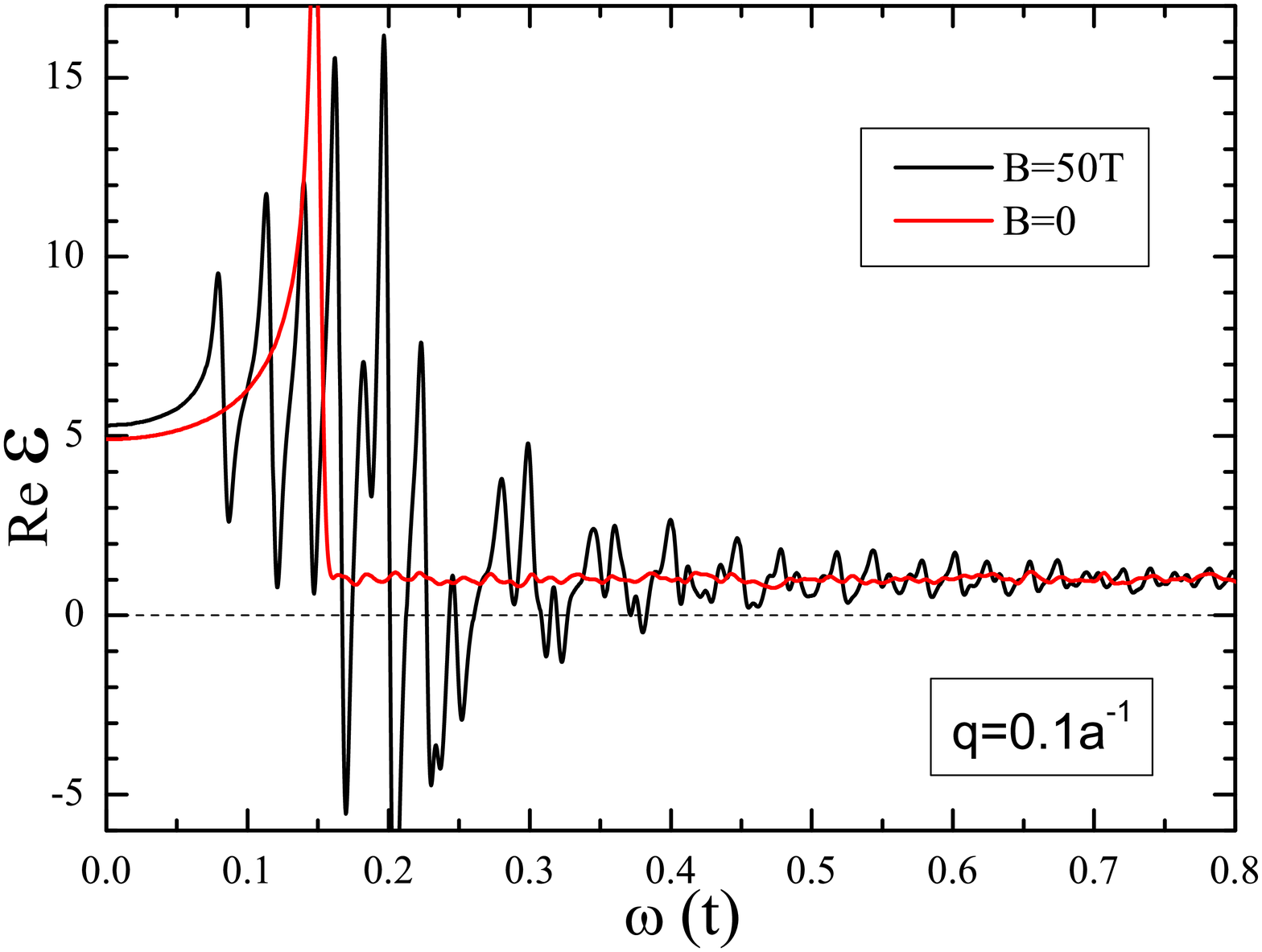}
\includegraphics[width=4.5cm]{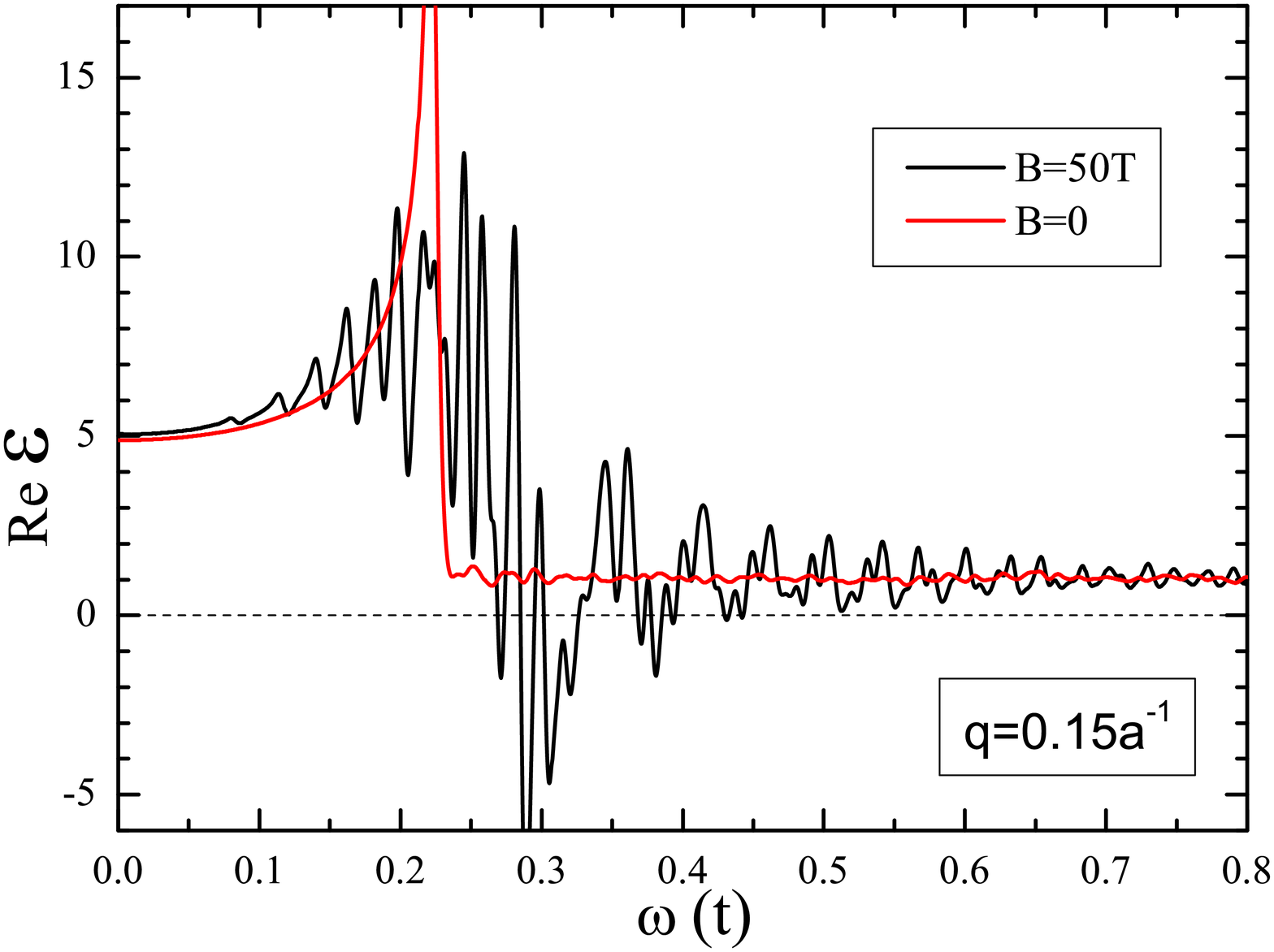}
} \mbox{
\includegraphics[width=4.5cm]{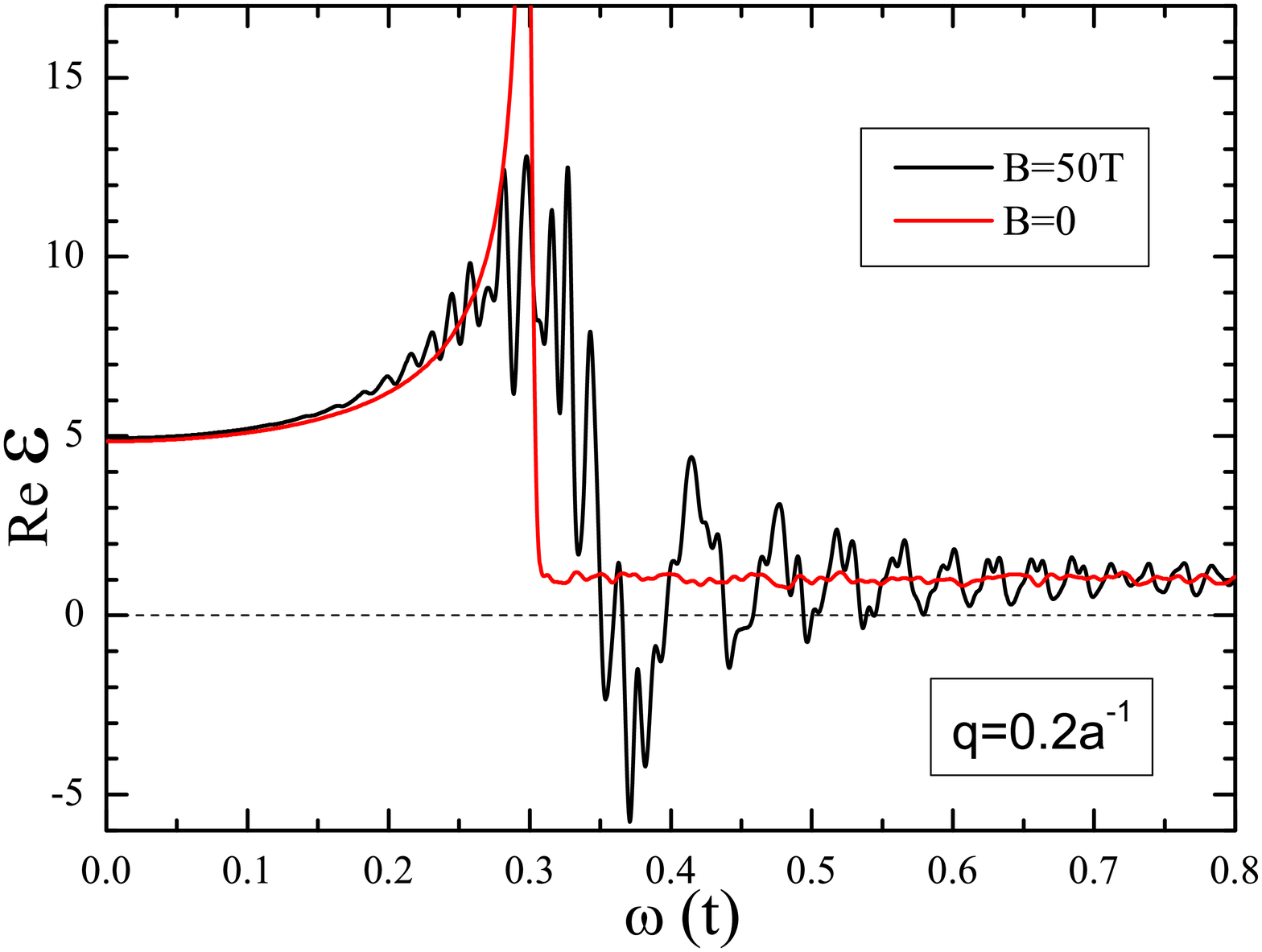}
}
\end{center}
\caption{$\mathrm{Re}~\protect\varepsilon(\mathbf{q},\protect\omega)$ for $%
B=0$ (red lines) and $B=50$T (black lines), for the values of the
wave-vectors used in Fig. \protect\ref{Fig:Loss}.}
\label{Fig:ReEpsilon}
\end{figure}

In the previous section we have discussed the excitation spectrum in the
absence of electron-electron interaction. In this section we include in the
problem the effect of long range Coulomb interaction. The polarization and
dielectric functions are calculated within the RPA, Eqs. (\ref{Eq:chi}) and (%
\ref{Eq:epsilon}). Within this framework, the existence of collective
excitations will be identified by the zeros of the dielectric function or
equivalently by the divergences of the loss function $-\mathrm{Im}%
[1/\epsilon (\mathbf{q},\omega )]$, which is proportional to
the spectrum measured by Electron Energy Loss Spectroscopy (EELS).\cite{EB08} In Fig. %
\ref{Fig:Loss} we show the loss function of graphene in a magnetic field, as
compared to the one at $B=0$. At $B=0$ the main structure is the broad peak
at $\omega \sim 2t$, associated to the $\pi $-plasmon.\cite{YRK11} When the
graphene layer is subjected to a strong perpendicular magnetic field, $%
\mathrm{Im}[1/\varepsilon (\mathbf{q},\omega )]$ presents a series of
prominent peaks at low frequencies, associated to collective modes in the
QHE regime, as it may be seen in Fig. \ref{Fig:Loss}(b). Similarly to the
single-particle case discussed in Sec. \ref{Sec:PHES}, the strength of the
peak is determined by the Coulomb matrix elements $V(q)|\mathcal{F}%
_{n,n^{\prime }}(\mathbf{q})|^{2}$, which depends strongly on the
wave-vector $q$. We emphasize that these modes can not be understood as a
simple many-body renormalization of the dispersionless inter-LL transitions
given in Eq. (\ref{Eq:PHDirac}), because only the low energy and long
wavelength modes have their non-interacting counterpart $\omega
_{n,n^{\prime }}$ associated to a specific single-particle electron-hole
transition with well defined indices $n$ and $n^{\prime }$. As we go to
higher energies and/or weaker magnetic fields, the relativistic LL
quantization of graphene leads to a so strong LL mixing that the collective
modes cannot be labeled any more in terms of single-particle excitations,%
\cite{RFG10,LS11} as in the case of a 2DEG with a quadratic dispersion and a
set of equidistant LLs.\cite{KH84}

In Fig. \ref{Fig:ReEpsilon} we compare the real part of the dielectric
function $\mathrm{Re}~\varepsilon(\mathbf{q},\omega)$ for zero and finite
magnetic field. The zeros of $\mathrm{Re}~\varepsilon(\mathbf{q},\omega)$
correspond to the frequencies of the undamped collective modes. Contrary to
the $B=0$ case, for which there is no collective modes for undoped graphene
at the RPA level, at $B\neq 0$ we observe a number of well defined zeros for 
$\mathrm{Re}~\varepsilon$, which correspond to coherent and long-lived
linear magneto-plasmons.\cite{RFG09} The divergence of $\mathrm{Im}\Pi(\mathbf{q}%
,\omega) $ at $\omega=v_Fq$ and the absence of backscattering in graphene
make that, as we increase the wave-vector $q$, the more coherent collective
modes are defined also for higher frequencies, namely around the threshold $%
\omega\approx v_Fq$, which is the frequency associated to the highest peaks
in Fig. \ref{Fig:Loss}(b). For even higher energies, the main contribution
to the modes of large frequencies are inter-LL transitions between well
separated LLs, with the subsequent reduction in the overlap in the
electron-hole wavefunctions. Therefore, the collective modes will suffer a
stronger Landau damping as higher frequencies are probed.

\subsection{Effect of disorder}

\begin{figure}[t]
\begin{center}
\includegraphics[width=7.3cm]{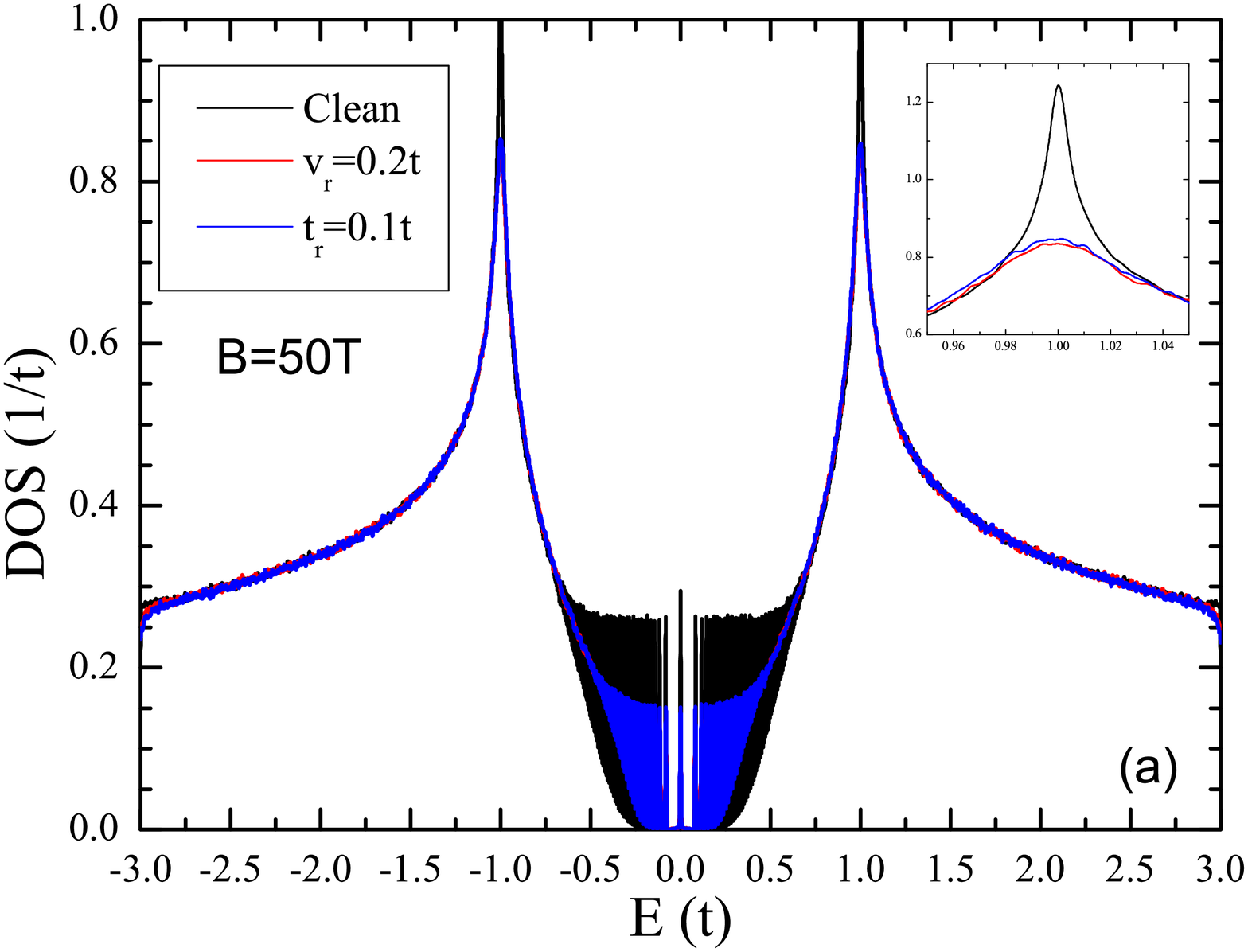} %
\includegraphics[width=7.3cm]{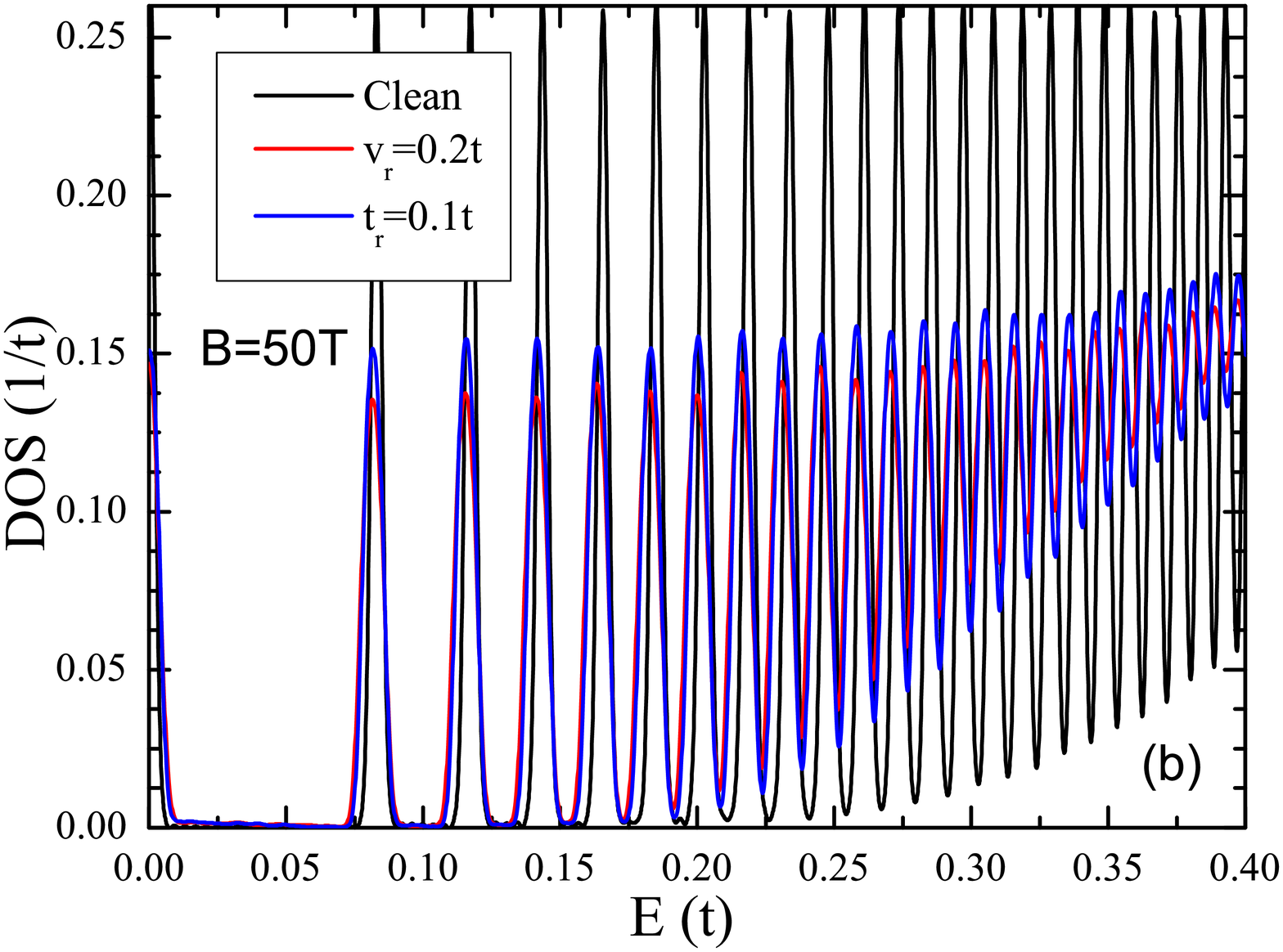}
\end{center}
\caption{(a) DOS of clean graphene (black lines) and of disordered graphene
with a random on-site potential (red lines) and with a random
renormalization of the hopping integrals (blue lines). The inset shows the smearing of the VHS peak due to disorder. (b) Zoom of the low
energy part of the spectrum. }
\label{Fig:DOSdisorder}
\end{figure}

Now we focus our attention on the effect of disorder on the DOS and on the
excitation spectrum of graphene in the QHE regime. In general, disorder
leads to a broadening of the LLs, with extended (delocalized) states near
the center of the original LL, and localized states in the tails. We
consider here two different kinds of disorder, namely random local change of
the on-site potentials $v_i$, which acts as a chemical potential shift for
the Dirac fermions, and random renormalization of the hopping amplitudes $%
t_{ij}$, due e. g. to changes of distances or angles between the carbon $p_z$
orbitals. They enter in the single-particle Hamiltonian as given in Eq. (\ref%
{Eq:Hamiltonian}). The effect of correlated long-range hopping
disorder has been shown to lead to a splitting of the $n=0$ LL,\cite%
{P09,PLM11} originated from the breaking of the sublattice and valley degeneracy.
Other kinds of disorder as vacancies create midgap states that
make the $n=0$ LL to remain robust, whereas the rest of LLs are smeared out
due to the effect of disorder.\cite{YRK10} In Fig. \ref{Fig:DOSdisorder} we
show the DOS of graphene in a perpendicular magnetic field of $B=50$T for
different kind of disorder, as compared to the clean case. We let the
on-site potential $v_i$ to be randomly distributed (independently on each
site $i$) between $-v_r$ and $+v_r$. On the other hand, the nearest-neighbor
hopping $t_{ij}$ is random and uniformly distributed (independently on sites 
$i,j$) between $t-t_r$ and $t+t_r$. At high energies, as we have seen in
Sec. \ref{Sec:DOS}, the DOS for this strength of the magnetic field is quite
similar to the DOS at $B=0$. Therefore, as in the zero field case,\cite%
{YRK10,YRK11} the main effect away the Dirac point is a smearing of the VHS
at $|\epsilon|=t$, as it is observed in the inset of Fig. \ref%
{Fig:DOSdisorder}(a).

\begin{figure*}[t]
\begin{center}
\mbox{
\includegraphics[width=6.cm]{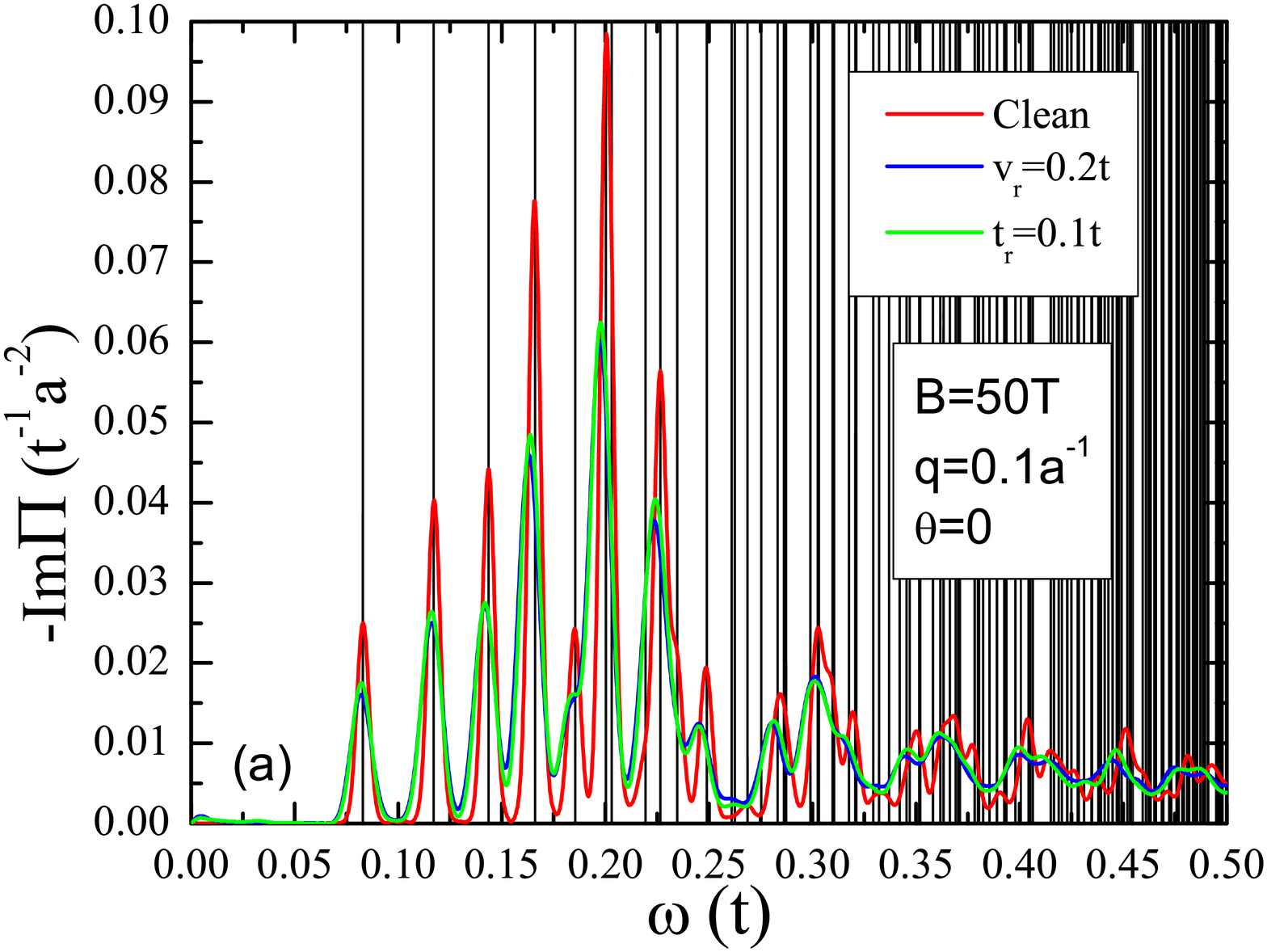}
\includegraphics[width=6.cm]{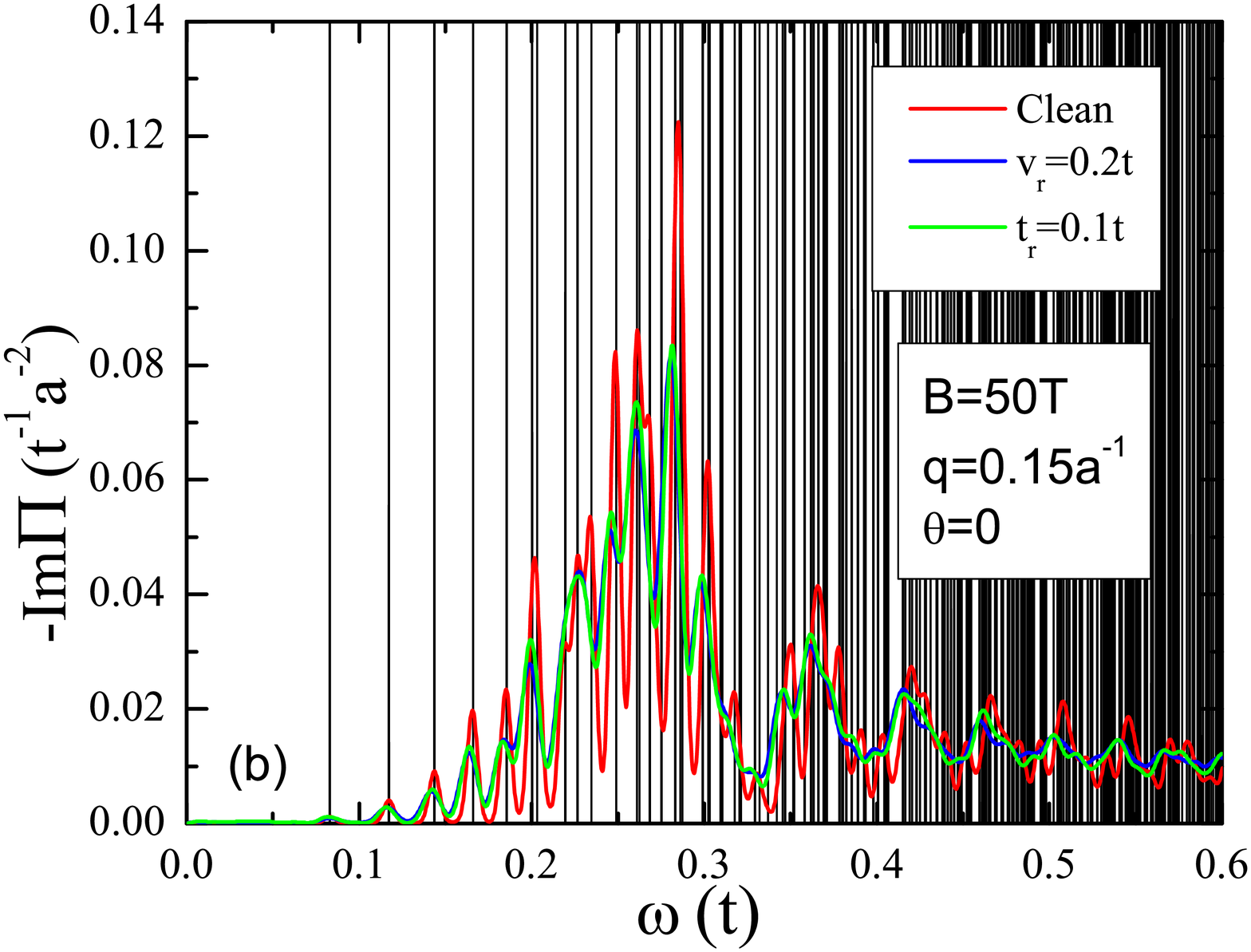}
\includegraphics[width=6.cm]{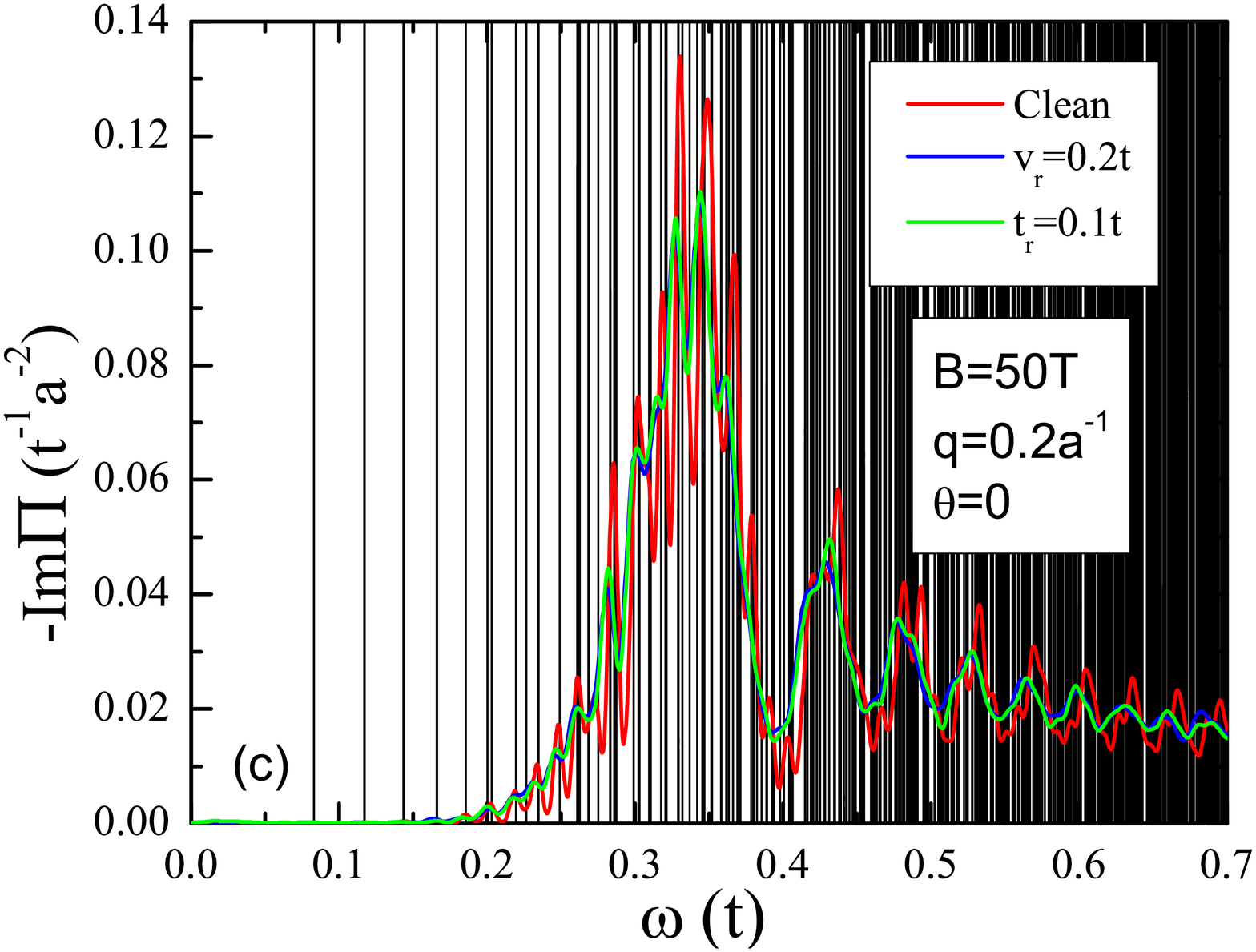}
} 
\mbox{
\includegraphics[width=6.cm]{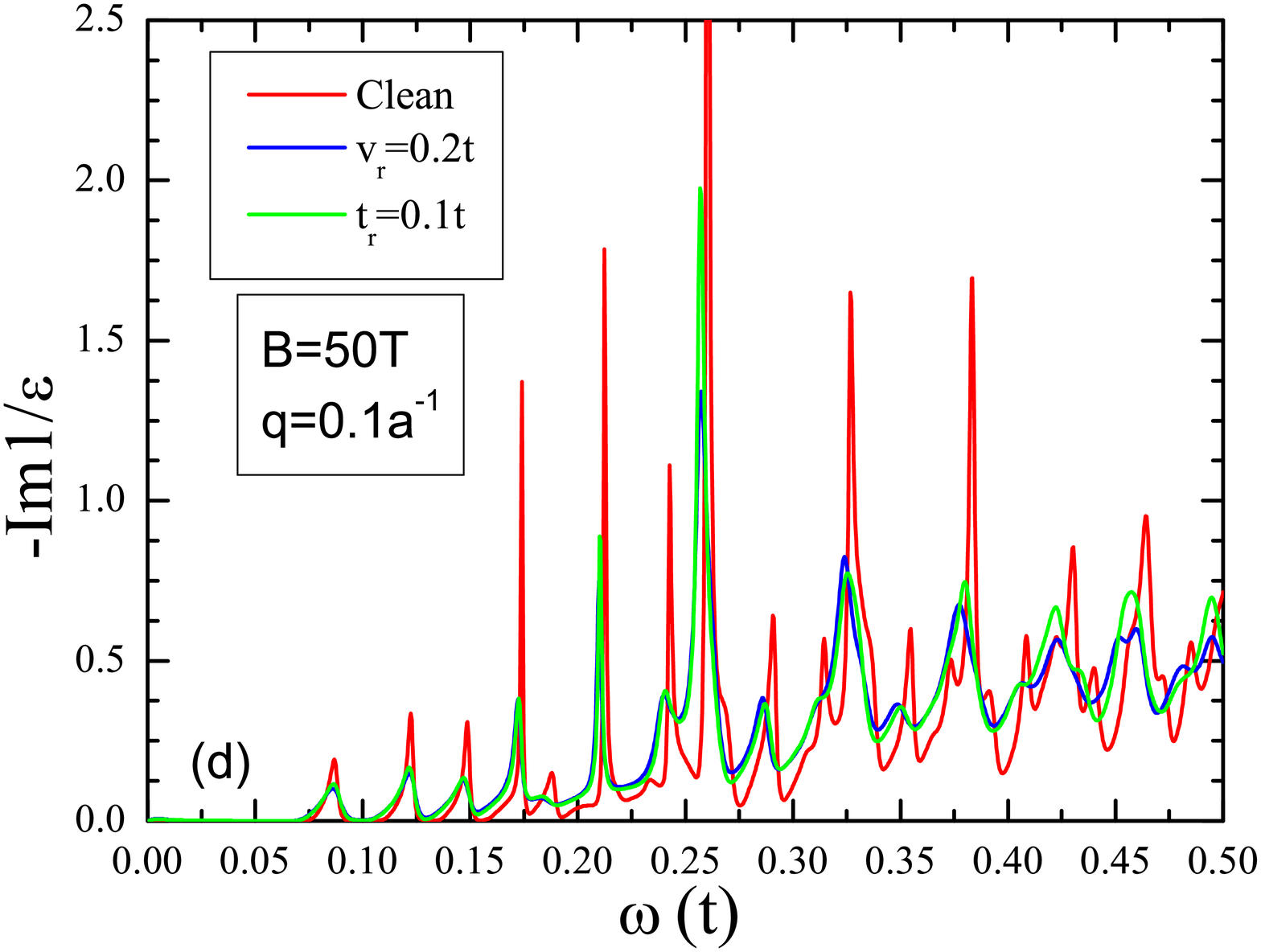}
\includegraphics[width=6.cm]{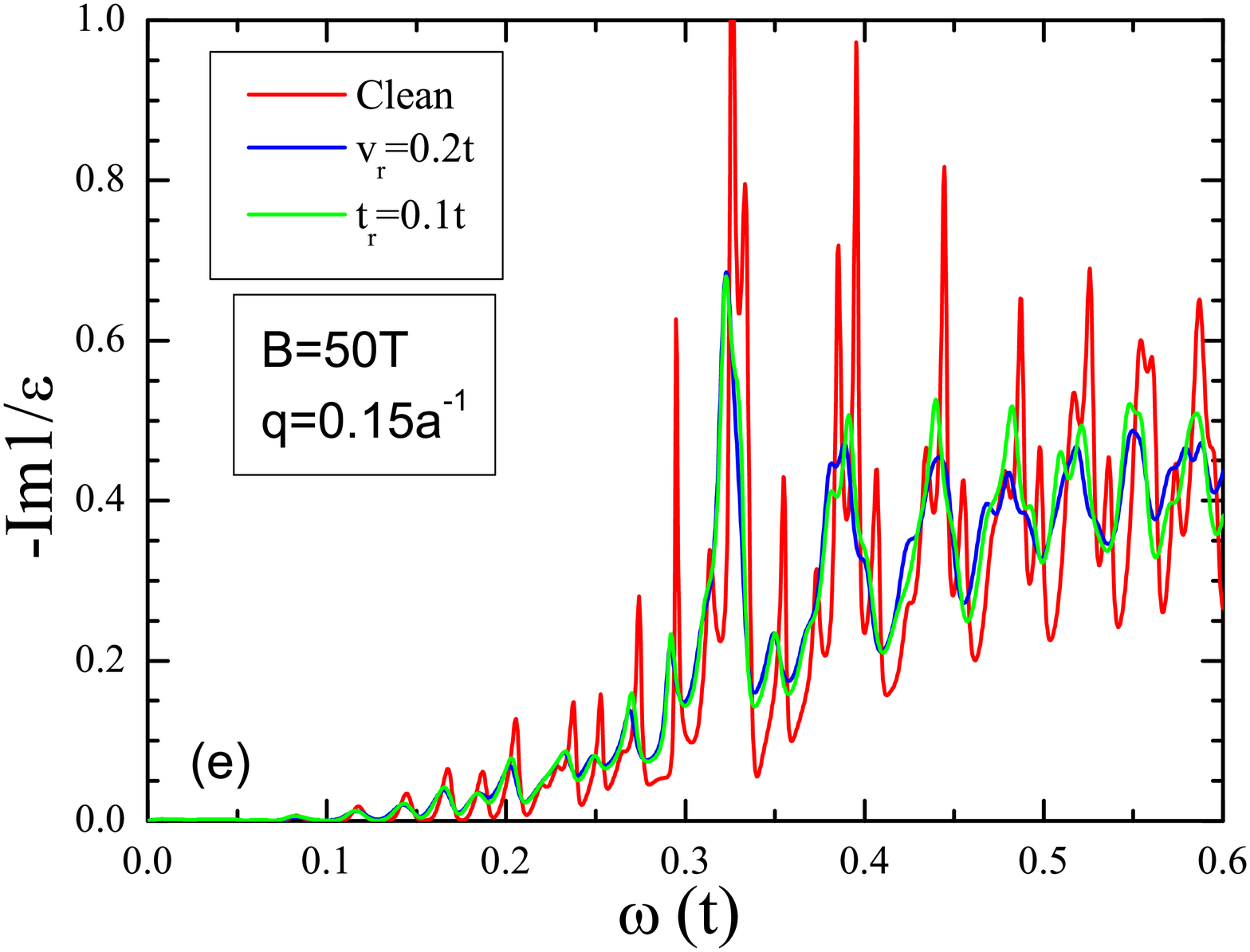}
\includegraphics[width=6.cm]{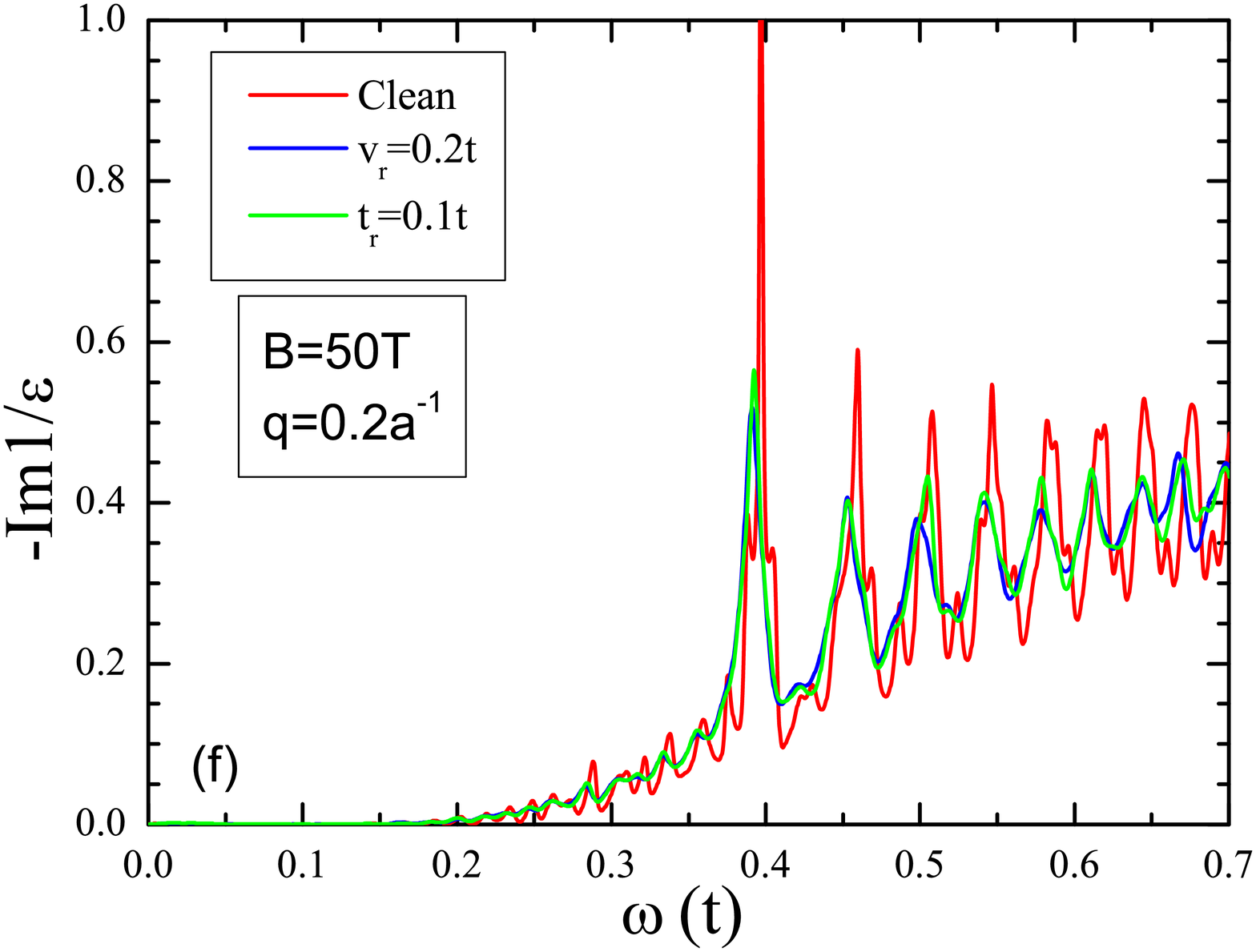}
} 
\mbox{
\includegraphics[width=6.cm]{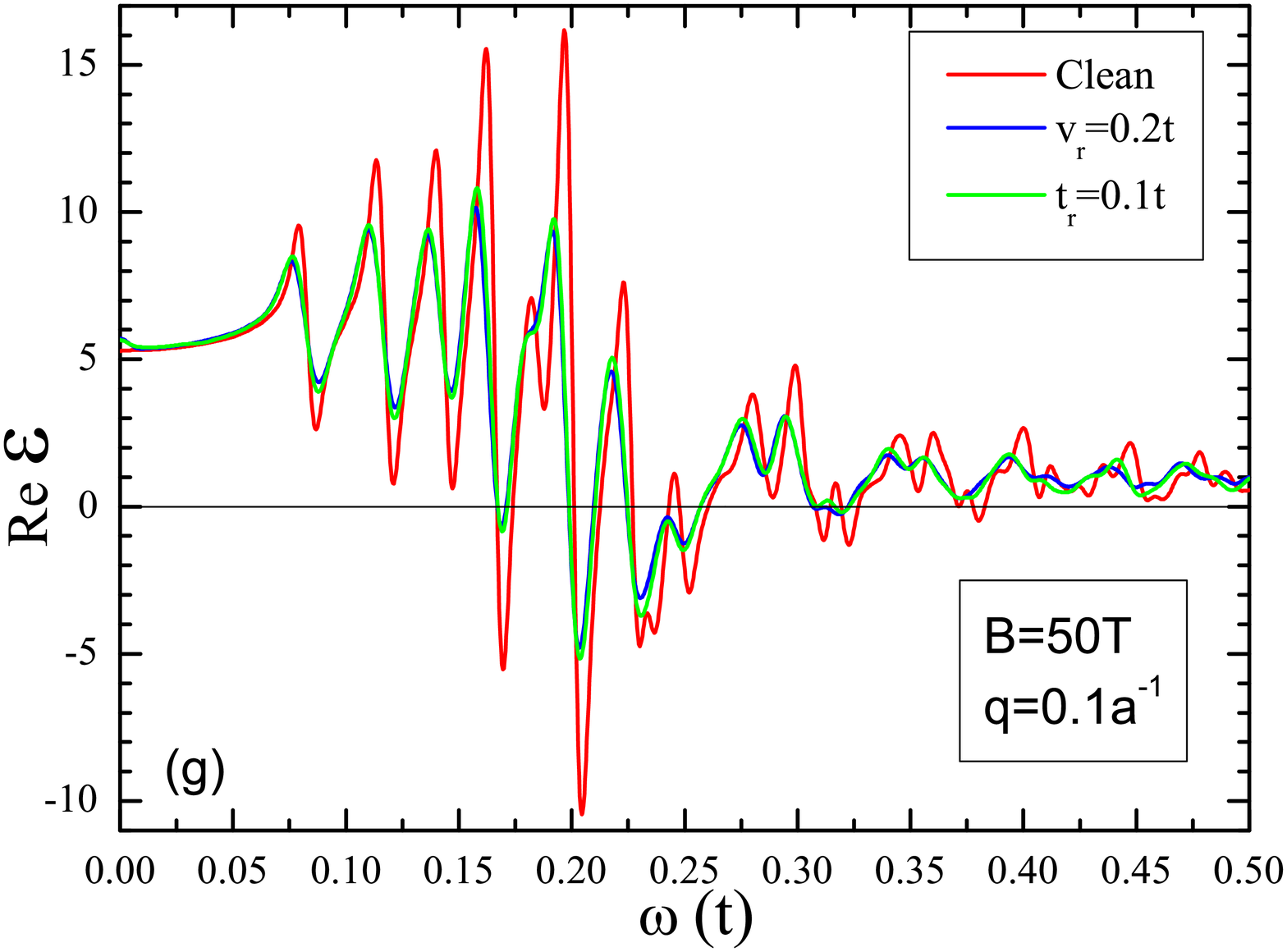}
\includegraphics[width=6.cm]{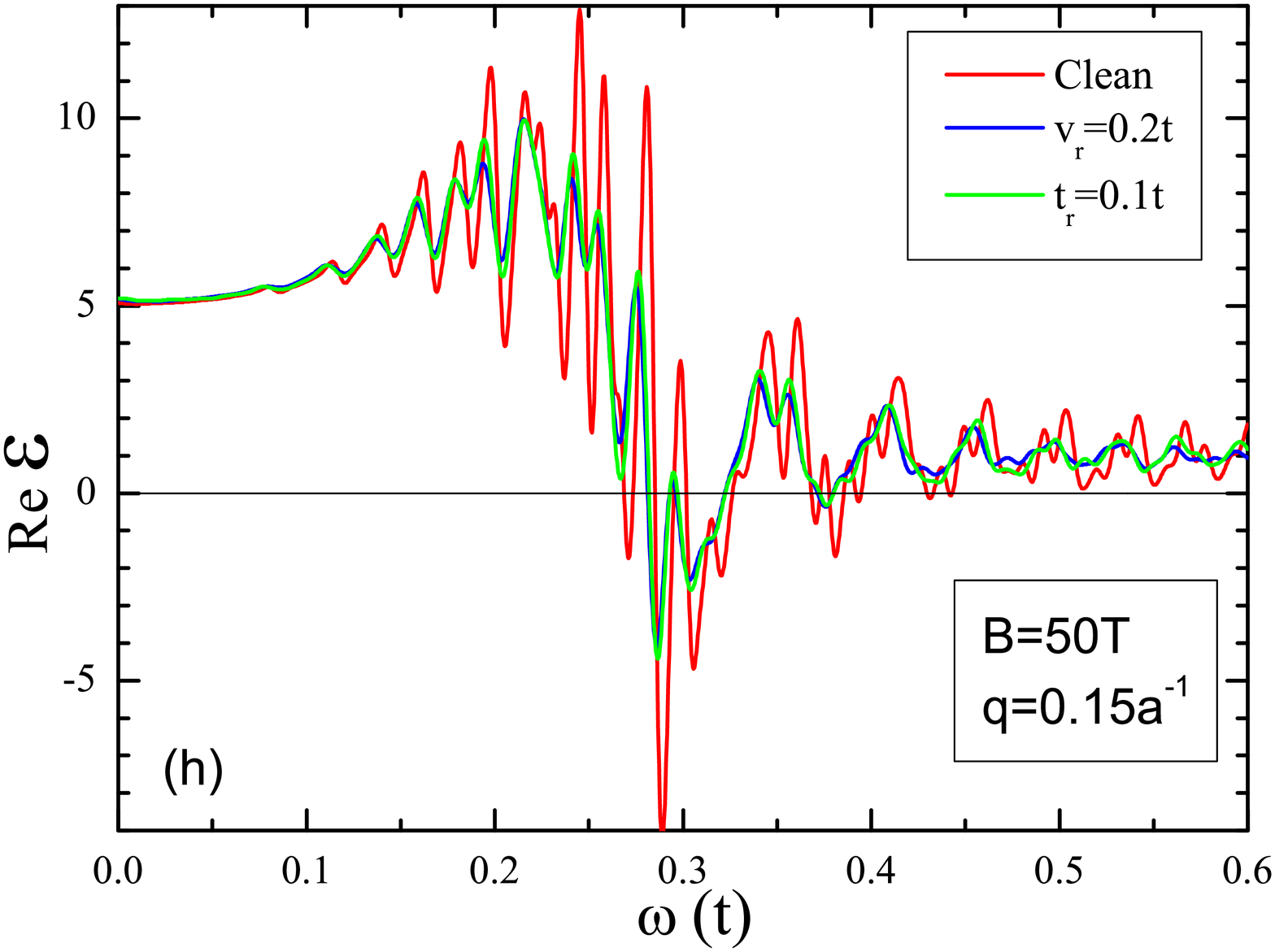}
\includegraphics[width=6.cm]{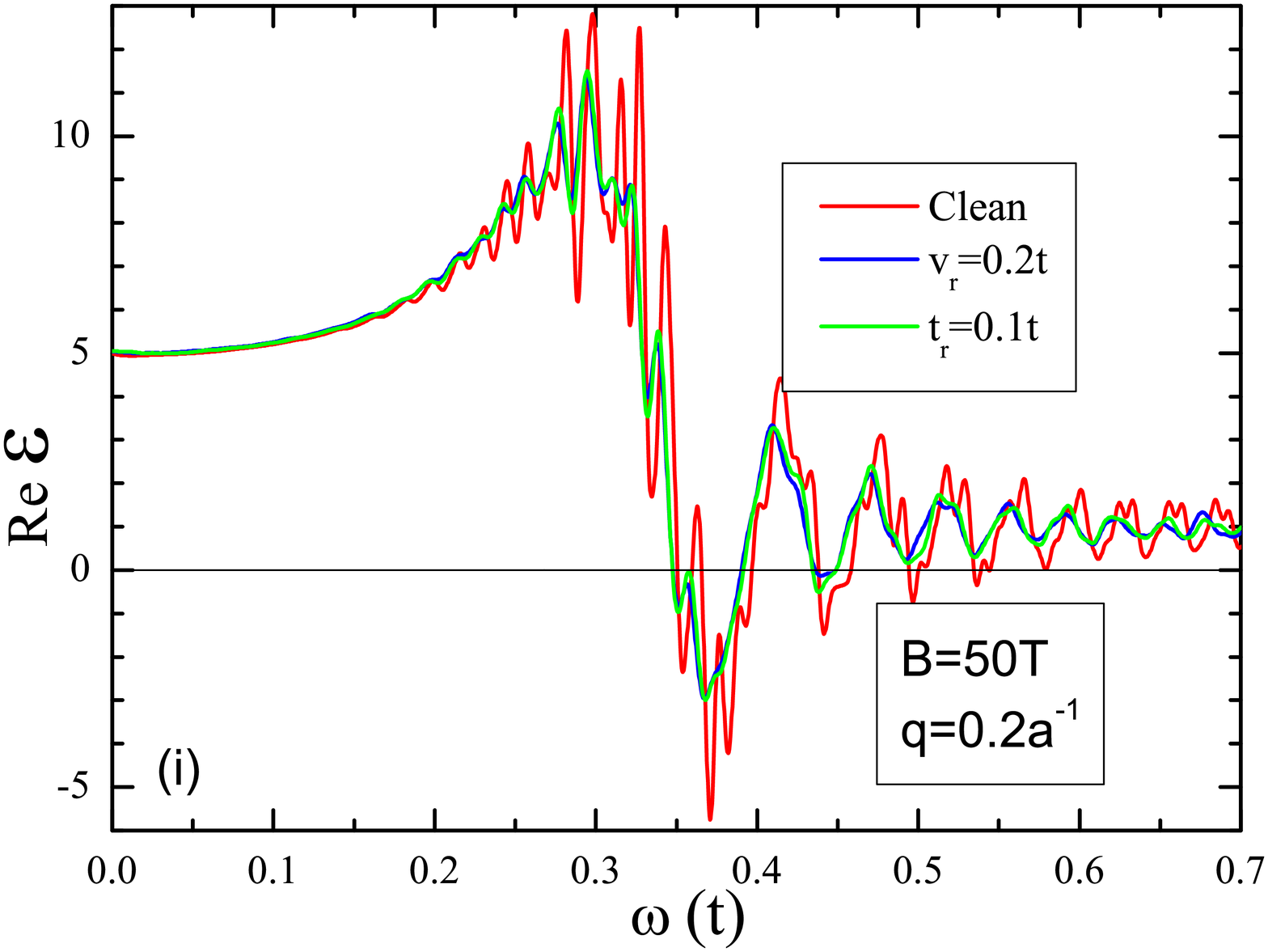}
}
\end{center}
\caption{(a)-(c) Non-interacting single-particle excitation spectrum of
graphene in a magnetic field of $B=50$T, as defined by $-\mathrm{Im}\Pi(%
\mathbf{q},\protect\omega)$, for different kinds of disorder and for
different values of the wave-vector $q$. The black vertical lines signal the
energy of the electron-hole processes defined by Eq. (\protect\ref%
{Eq:PHDirac}). (d)-(f) Loss function $-\mathrm{Im}[1/\protect\varepsilon(%
\mathbf{q},\protect\omega)]$ in the RPA. (g)-(i) Real part of the dielectric
function $\mathrm{Re}~\protect\varepsilon(\mathbf{q},\protect\omega)$ in the
RPA. }
\label{Fig:PHESdisorder}
\end{figure*}

Well defined LLs occur around the Dirac point, and the effect of disorder on
the peaks is observed in Fig. \ref{Fig:DOSdisorder}, where we show a zoom of
the low energy part of the DOS around $\epsilon =0$. Both kinds of disorder
(random on-site potential and random hopping) leads to a similar effect,
producing a broadening of the LLs. We can also observe,
especially for the highest LLs shown in Fig. \ref{Fig:DOSdisorder}(b), a tiny
but appreciable redshift of the position of the center of the LLs with
respect to its original position, in agreement with previous works.\cite{SA98,PLM11} The full self-consistent Born approximation
calculations for graphene with unitary scatterers of Ref. \onlinecite{PGC06}
leaded to a rather significant change in the position of the LLs towards
higher energies. However, the exact transfer matrix and diagonalization
calculations of Ref. \onlinecite{GJC08} found only a small shift of the LL
position for very strong disorder.

We now study the effect of disorder on the PHES. In Fig. \ref%
{Fig:PHESdisorder}(a)-(c) we show $-\mathrm{Im}\Pi(\mathbf{q},\omega)$ for
different values of $q$, and for different kinds of disorder. First we
notice a smearing of the resonance peaks associated to the LL broadening due
to disorder. Whereas for low energies the position of the resonance peaks of
disordered graphene coincides with the position for the clean case, we
observe a redshift of the resonance peaks as we consider inter-LL
transitions of higher energies. This is due to the change in the position of
the high energy LLs of disordered graphene with respect to the original LLs,
as we have discussed previously [see Fig. \ref{Fig:DOSdisorder}(b)].

The presence of disorder will also affect the dispersion relation and
coherence of the collective modes when the effect of electron-electron
interaction is included. In particular, the effect of a short range disorder
on graphene in a magnetic field can lead, due to the possibility of
inter-valley processes associated to the breakdown of sublattice symmetry,
to the localization of some collective modes on the impurity.\cite{FRD11} In
Fig. \ref{Fig:PHESdisorder}(d)-(f) we show the effect of a random on-site
potential or a random hopping renormalization on the loss function. First,
we can observe an important attenuation of the intensity peaks due to
disorder. Second, as we have discussed above, the position of the peaks of
disordered and clean graphene coincides at low energies, but not at high
energies, where the resonance peaks of the loss function of disordered
graphene are shifted with respect to clean graphene. Although we obtain a
similar renormalization of the spectrum for the two kinds of disorder
considered here, we reiterate that this effect is highly dependent on the
type disorder considered, as well as the theoretical method used to obtain
the spectrum.

Finally, we mention that the disorder LL broadening leads to an
amplification of the LL mixing discussed above. As a consequence, some
collective modes which are undamped for clean graphene, start to be Landau
damped due to the effect of disorder. Indeed, in Fig. \ref{Fig:PHESdisorder}%
(g)-(i) we can see how Eq. (\ref{Eq:Plasmons}), which is the condition for
the existence of coherent collective modes, is fulfilled more times for the
clean case than for the disordered membranes, for which the collective modes
are more highly damped.

\subsection{Effect of temperature: thermally activated electron-hole
transitions}

\begin{figure}[t]
\begin{center}
\mbox{
\includegraphics[width=4.3cm]{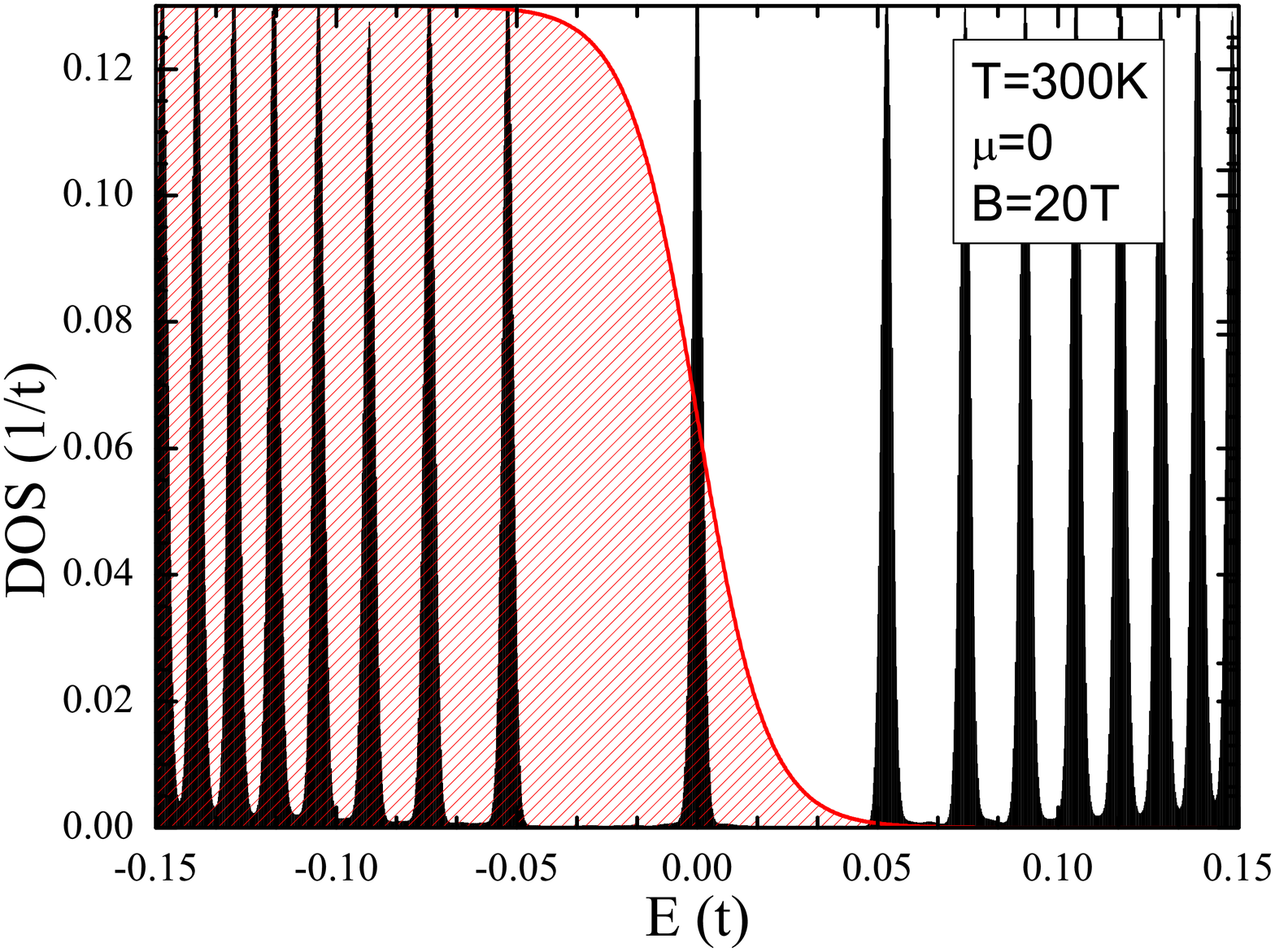}
\includegraphics[width=4.3cm]{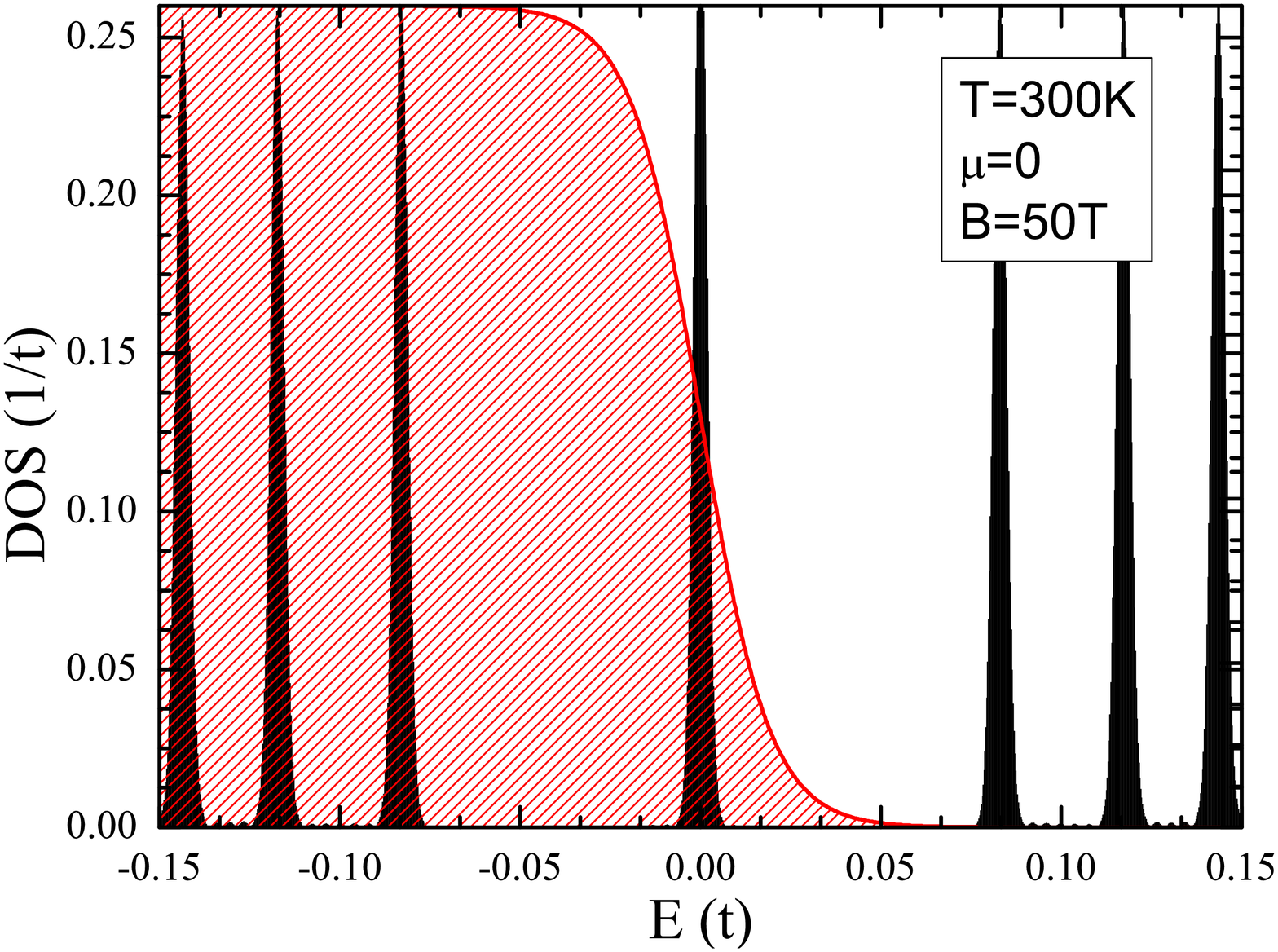}
} 
\mbox{
\includegraphics[width=4.3cm]{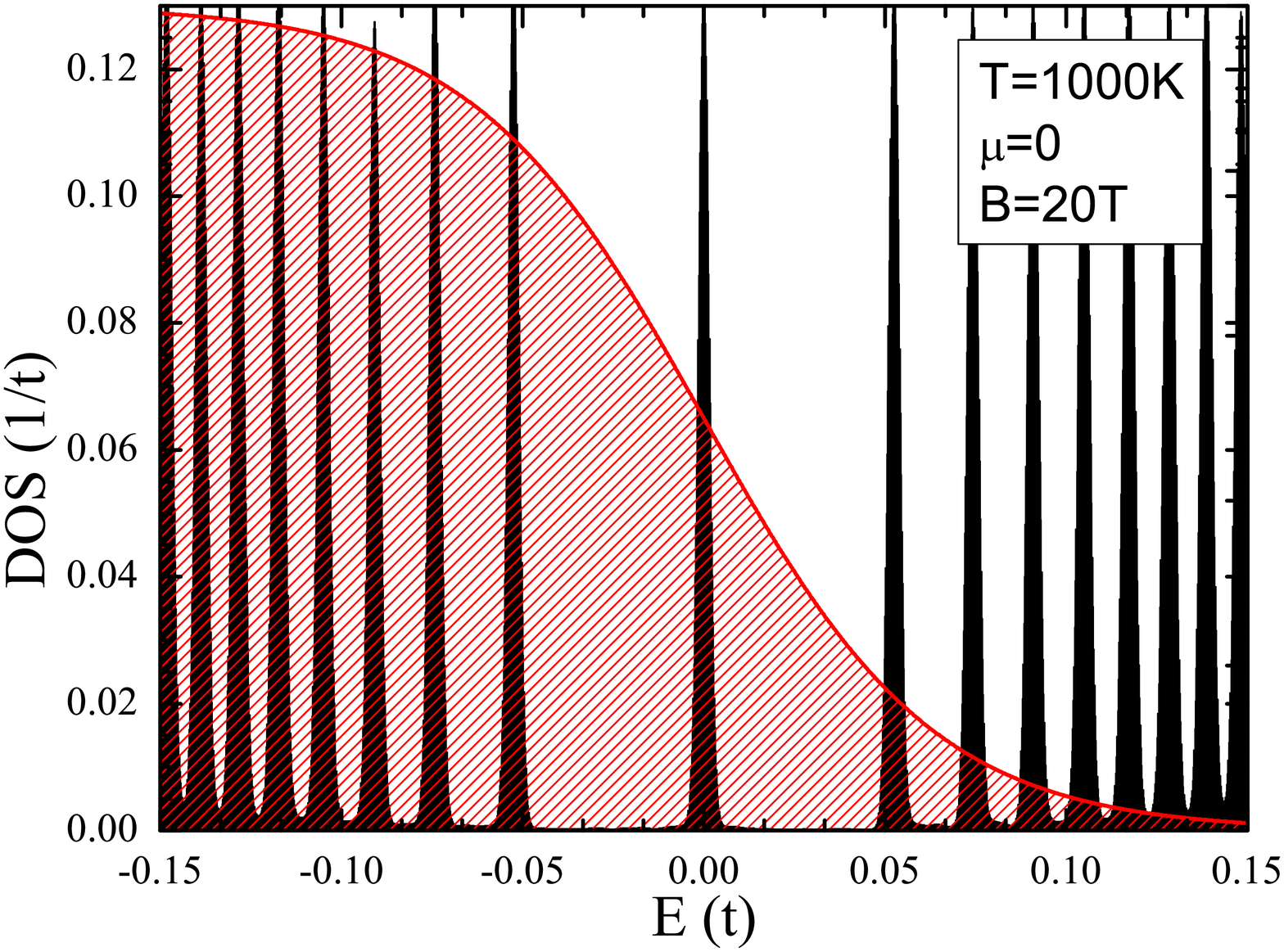}
\includegraphics[width=4.3cm]{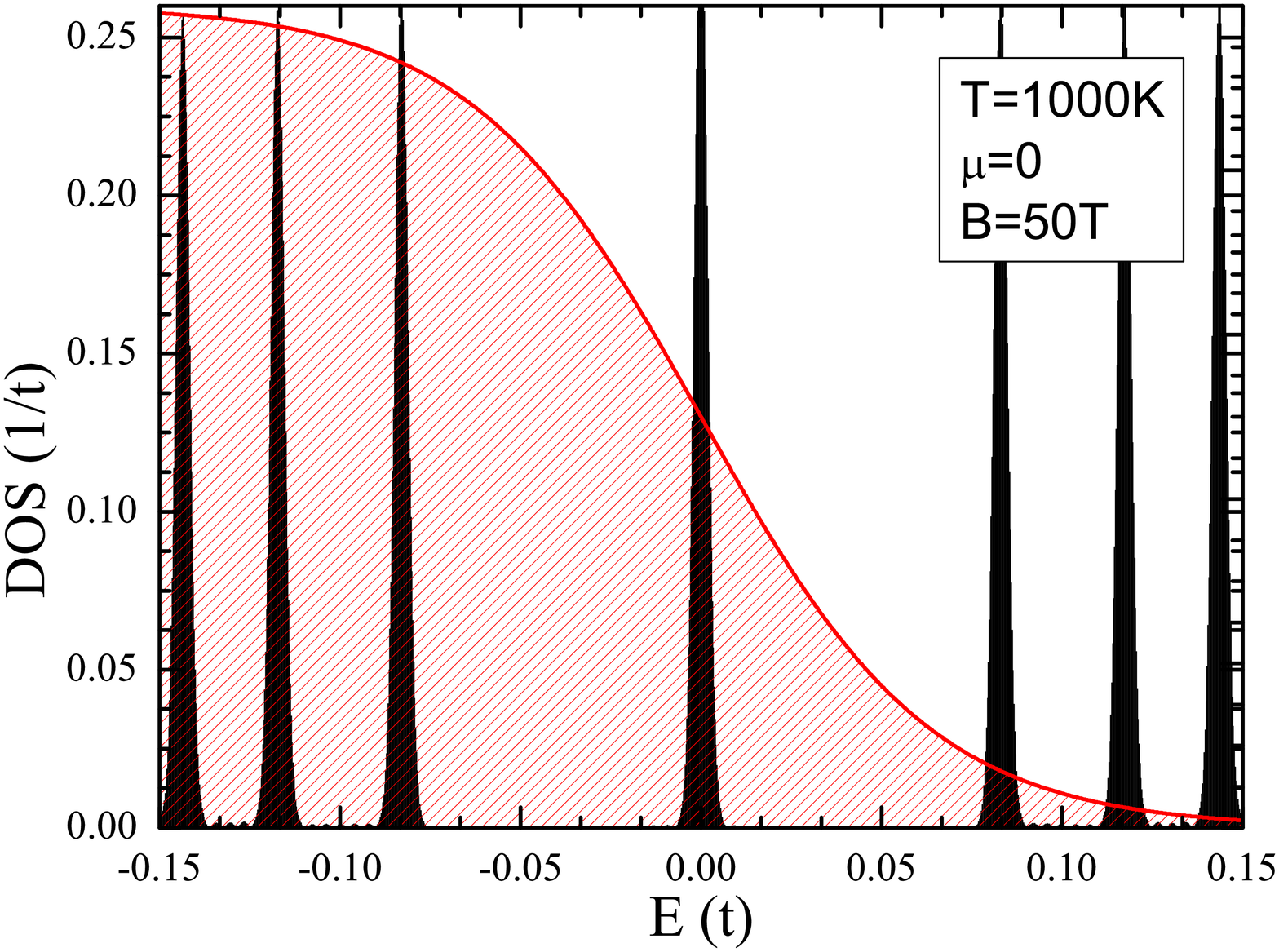}
}
\end{center}
\caption{DOS of clean graphene for different values of temperature $T$ and
magnetic field $B$. The shaded area is a sketch of the Fermi-Dirac
distribution function for each case.}
\label{Fig:Fermi-Dirac}
\end{figure}

\begin{figure*}[t]
\begin{center}
\mbox{
\includegraphics[width=6.3cm]{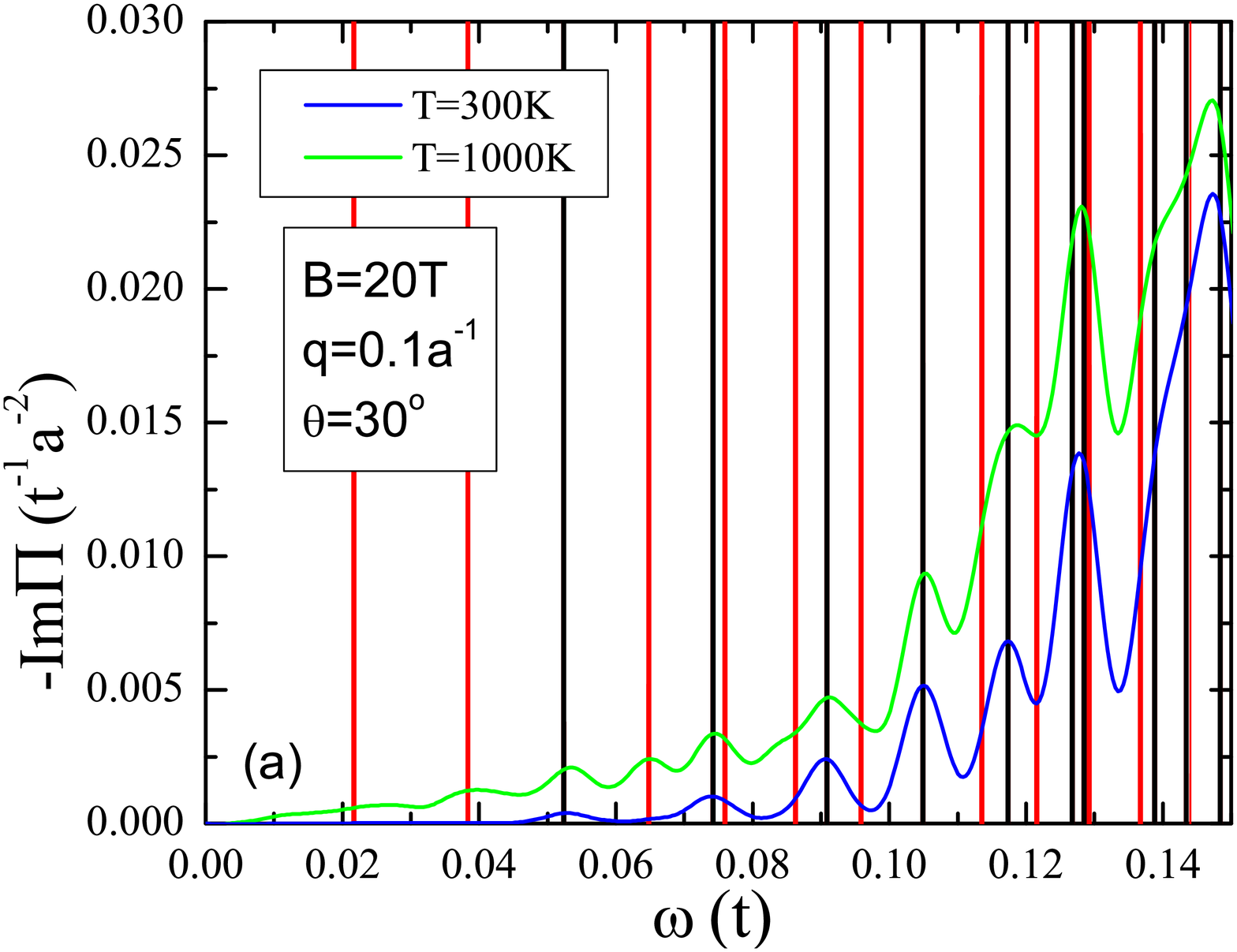}
\includegraphics[width=6.3cm]{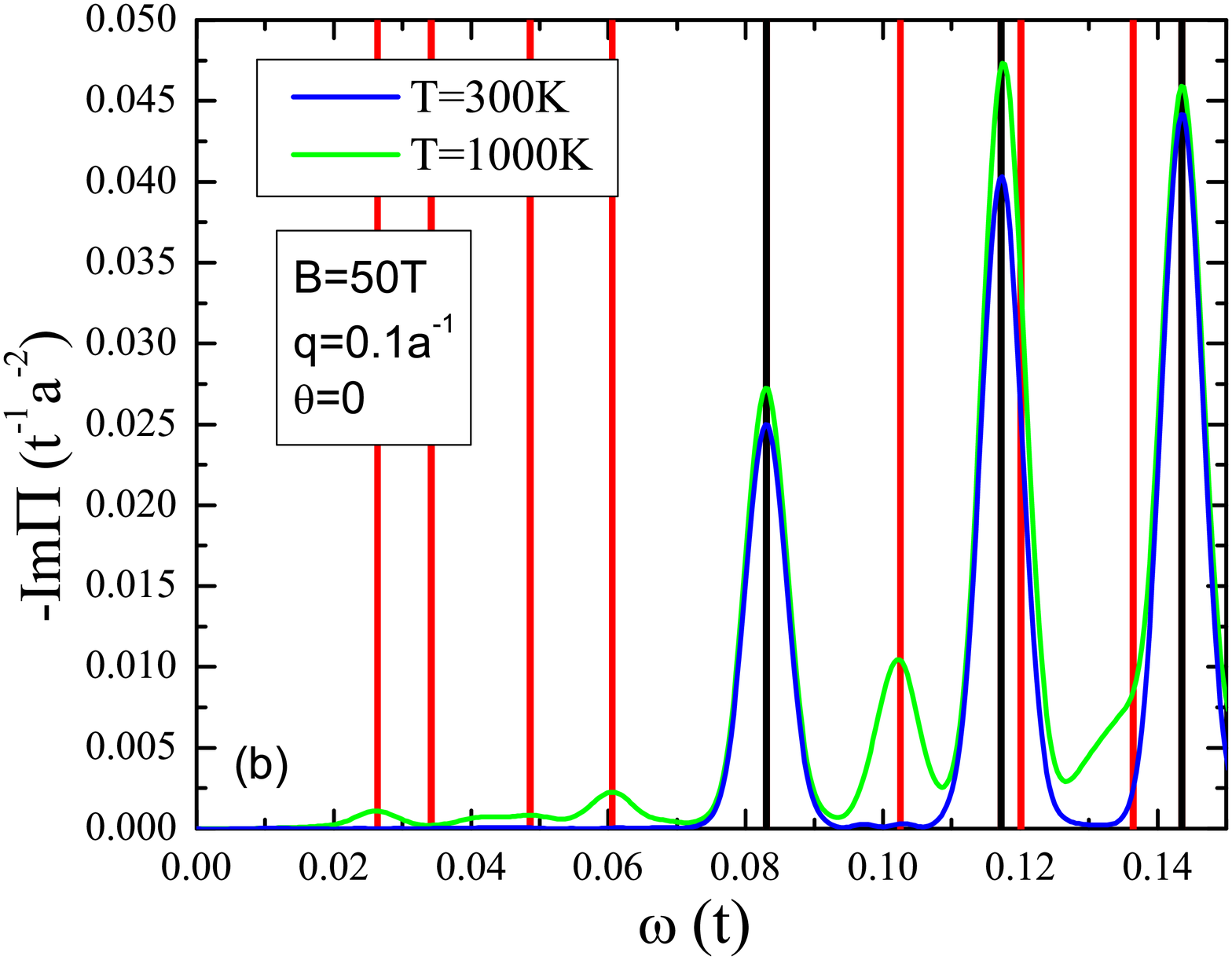}
} 
\mbox{
\includegraphics[width=6.3cm]{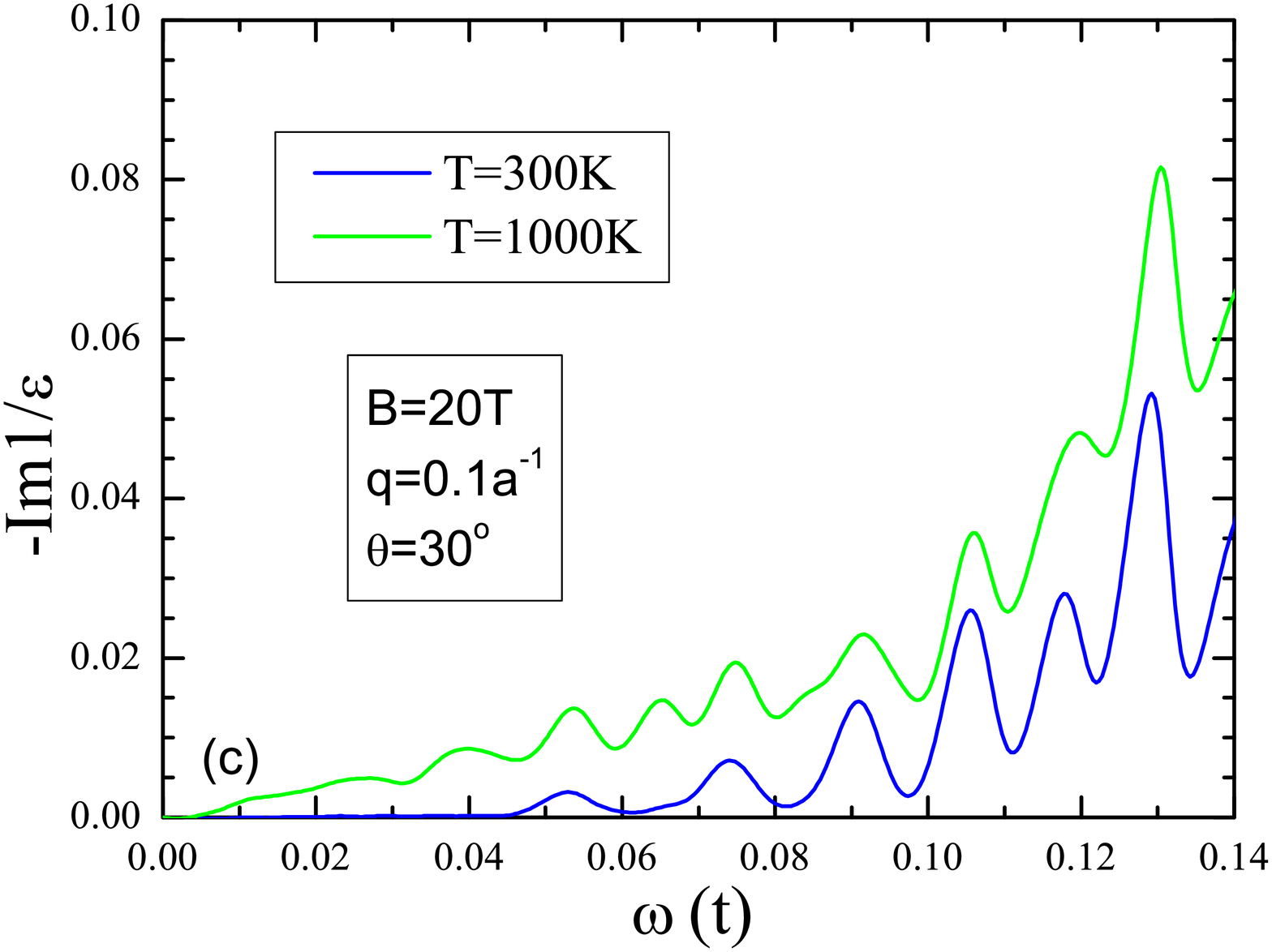}
\includegraphics[width=6.3cm]{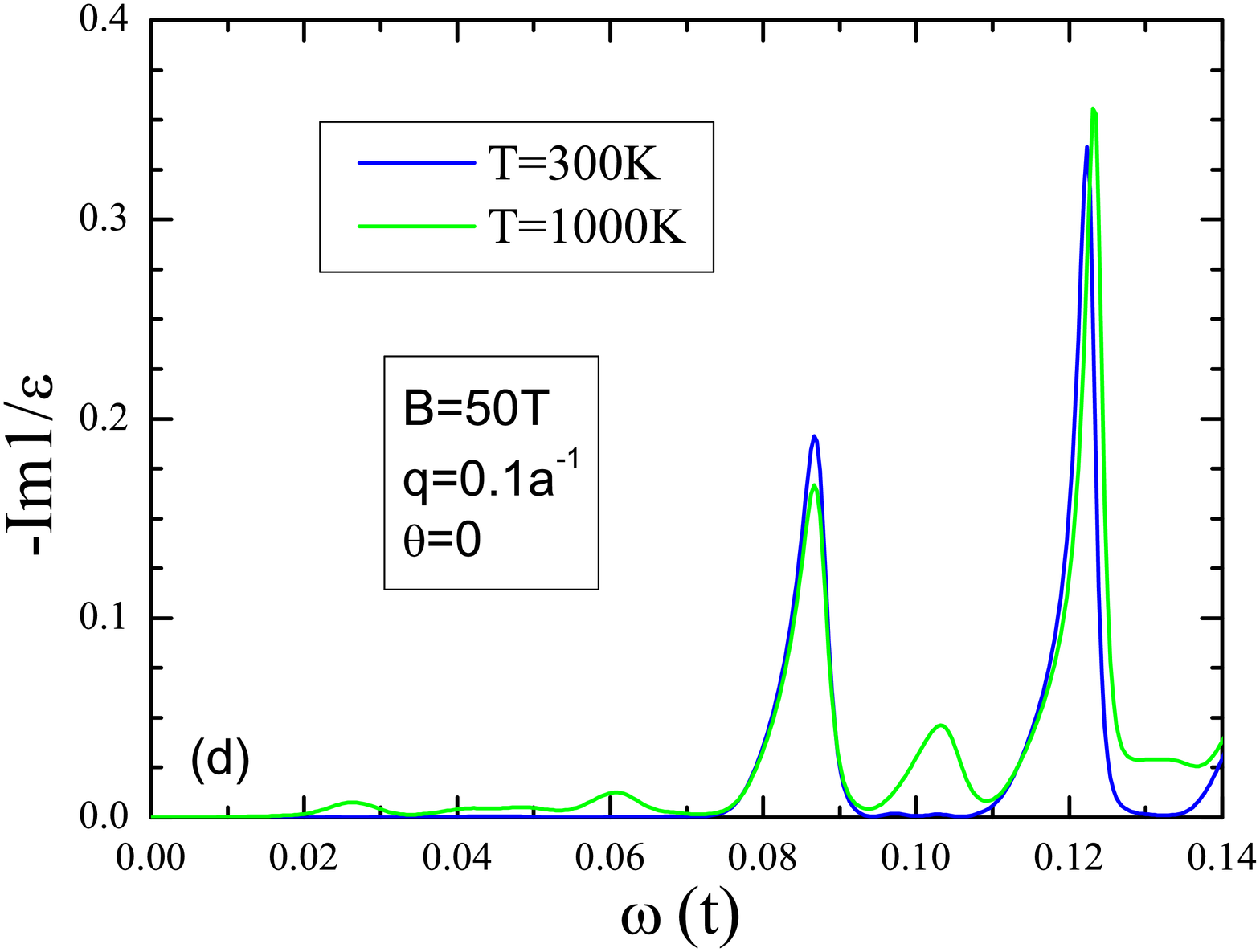}
}
\end{center}
\caption{(a)-(b) $-\mathrm{Im}\Pi(\mathbf{q},\protect\omega)$ for two
different values of $T$ and $B$. The black vertical lines denote the
position of the inter-LL transitions given by Eq. (\protect\ref{Eq:PHDirac})
which are possible at $T=0$. The red vertical lines correspond to the energy
of thermally activated inter-LL transitions. (c)-(d) Loss function $-\mathrm{%
Im}[1/\protect\varepsilon(\mathbf{q},\protect\omega)]$ for the same values
of $T$ and $B$. }
\label{Fig:Temp}
\end{figure*}

In this subsection we discuss the effect of temperature on the excitation
spectrum, which enters in our calculations trough the Fermi-Dirac
distribution function Eq. (\ref{Eq:Fermi-Dirac}). Temperature can activate
additional inter-LL transitions due to free thermally induced electrons and
holes in the sample. In Fig. \ref{Fig:Fermi-Dirac} we sketch the LLs of
undoped clean graphene that are thermally activated, for two different
strengths of the magnetic field. For this, the corresponding Fermi-Dirac
distribution function is sketched on each plot, indicating the LLs that can
be partially populated with electrons (holes) in the conduction (valence)
band. Notice that at $T=0$ and for $\mu=0$, $n_F(\epsilon)$ is just a step
function that traverses the $n=0$ LL. Of course, the number of activated
LLs, which are those crossed by the tail of $n_F(\epsilon)$, grows as we
increase the temperature and/or as we decrease the magnetic field. The
population effect due to the thermal excitations of carriers has been
observed by far infrared transmission experiments.\cite{OP08}

In Fig. \ref{Fig:Temp} we show the single-particle polarization and the loss
function for two values of temperature and magnetic field. At room
temperature and for the rather strong magnetic fields considered in our
calculation, the allowed electron-hole transitions are the same as in the
zero temperature limit (see the top panels of Fig. \ref{Fig:Fermi-Dirac}).
Therefore, the peaks of $\mathrm{Im}\Pi$ for $T=300$~K are centered at the
frequencies of inter-LL transitions marked by the black vertical lines,
which accounts only for the usual inter-band transitions across the Dirac
point. For a considerable higher temperature of $T=10^3$~K there are
additional electron-hole transitions (some of them intra-band processes,
especially important at low frequencies) which are now allowed due to the
effect of temperature, as marked by the red vertical lines in Fig. \ref%
{Fig:Temp}(a)-(b). These thermally activated inter-LL transitions at high
temperatures contribute to the additional spectral weight of the PHES of
Fig. \ref{Fig:Temp}(a)-(b). Finally, in Fig. \ref{Fig:Temp}(c)-(d) we show
the loss functions corresponding to the magnetic fields and temperatures
discussed above. As we have discussed above, the peaks of $\mathrm{Im}%
~1/\varepsilon$ correspond to the position of collective excitations. Here
we find, in agreement with previous tight-binding and band-like matrix
numerical methods,\cite{WL11} a weak but appreciable renormalization of the
collective mode peak position as a function of temperature. This temperature
dependence of the collective mode is easily noticed by comparing the red and
blue peaks at $\omega\approx 0.12t$ of Fig. \ref{Fig:Temp}(d).

\section{Conclusions}

In conclusion, we have studied the excitation spectrum of a graphene layer
in the presence of a strong magnetic field, using a full $\pi$-band
tight-binding model. The magnetic field has been introduced by means of a
Peierls substitution, and the effect of long range Coulomb interaction has
been considered within the RPA. For realistic values of the magnetic field,
the LL quantization leads to well defined LLs around the Dirac point,
whereas the DOS at higher energies is rater similar to the one at zero
field. However, we have shown that in the ultra-high magnetic field limit,%
\cite{SP11} for which the magnetic length is comparable to the lattice
spacing, the LL quantization around the Van Hove singularity is highly
nontrivial, with two different sets of LLs that merge at the saddle point.

Our results for the polarization function shows that, at high energies, the
PHES is dominated, as in the $B=0$ case,\cite{YRK11} by the $\pi$-plasmon,
which is associated to the enhanced DOS at the VHS of the $\pi$ bands. The
low energy part of the spectrum is however completely different to its zero
field counterpart. The relativistic LL quantization of the spectrum into
non-equidistant LLs lead to a peculiar excitation spectrum with a strong
modulation of the spectral weight, which can be understood in terms of the
node structure of the electron-hole wavefunction overlap.\cite{RGF10}
Furthermore, we have shown that the presence of disorder in the sample lead
to a smearing of the resonance peaks of the loss function, and to an
enhancement of the Landau damping of the collective modes. Finally, we have studied the effect of temperature on the spectrum, and shown that it can activate additional inter-LL transitions, effect
which is especially relevant at low magnetic fields.

\section{Acknowledgement}

The support by the Stichting Fundamenteel Onderzoek der Materie (FOM) and
the Netherlands National Computing Facilities foundation (NCF) are
acknowledged. We thank the EU-India FP-7 collaboration under MONAMI and the
grant CONSOLIDER CSD2007-00010.

\bibliographystyle{apsrev4-1}
\bibliography{BibliogrGrafeno}

\end{document}